\documentclass[sn-basic, Numbered]{sn-jnl}

\usepackage{graphicx}%
\usepackage{multirow}%
\usepackage{amsmath,amssymb,amsfonts}%
\usepackage{amsthm}%
\usepackage{mathrsfs}%
\usepackage{xcolor}%
\usepackage{textcomp}%
\usepackage{manyfoot}%
\usepackage{booktabs}%
\usepackage{algorithm}%
\usepackage{algorithmicx}%
\usepackage{algpseudocode}%
\usepackage{listings}%
\usepackage{subcaption}
\usepackage{soul}

\theoremstyle{thmstyleone}%
\theoremstyle{thmstyletwo}%

\theoremstyle{thmstylethree}%

\begin{document}

\title[Capacity Planning and Scheduling for Jobs with Uncertainty in both Resource Usage and Duration]{Capacity Planning and Scheduling for Jobs with Uncertainty in Resource Usage and Duration}


\author[1]{\fnm{Sunandita} \sur{Patra}}
\author[2]{\fnm{Mehtab} \sur{Pathan}}
\author[3]{\fnm{Mahmoud} \sur{Mahfouz}}
\author[3]{\fnm{Parisa} \sur{Zehtabi}}
\author[2]{\fnm{Wided} \sur{Ouaja}}
\author[3]{\fnm{Daniele} \sur{Magazzeni}}
\author[1]{\fnm{Manuela} \sur{Veloso}}

\affil[1]{\orgdiv{AI Research}, \orgname{J.P. Morgan}, \city{New York},  \country{USA}}
\affil[2]{\orgdiv{CIB Athena Applied Intelligence}, \orgname{J.P. Morgan}, \city{London},  \country{UK}}
\affil[3]{\orgdiv{AI Research}, \orgname{J.P. Morgan}, \city{London},  \country{UK}}

\affil[]{\colorbox{yellow}{\texttt{Please cite as follows}}}
\affil[]{\texttt{\colorbox{yellow}{Sunandita Patra, Mehtab Pathan, Mahmoud Mahfouz, Parisa} \colorbox{yellow}{Zehtabi, Wided Ouaja, Daniele Magazzeni, and Manuela Veloso.} \colorbox{yellow}{ "Capacity planning and scheduling for jobs with uncertainty in} \colorbox{yellow}{resource usage and duration." The Journal of Supercomputing} \colorbox{yellow}{80, no. 15 (2024): 22428-22461}}}

\abstract{

    Organizations around the world schedule jobs (programs) regularly to perform various tasks dictated by their end users. With the major movement towards using a cloud computing infrastructure, our organization follows a hybrid approach with both cloud and on-prem servers. The objective of this work is to perform capacity planning, i.e., estimate resource requirements, and job scheduling for on-prem grid computing environments. A key contribution of our approach is handling uncertainty in \textit{both} resource usage and duration of the jobs, a critical aspect in the finance industry where stochastic market conditions significantly influence job characteristics.  For capacity planning and scheduling, we simultaneously balance two conflicting objectives: (a) minimize resource usage, and (b) provide high quality-of-service to the end users by completing jobs by their requested deadlines. We propose approximate approaches using deterministic estimators and pair sampling-based constraint programming. Our best approach (pair sampling-based) achieves improved peak reduction in resource usage compared to manual scheduling without compromising on the quality-of-service.
    
}

\keywords{capacity planning, stochastic resource usage, stochastic job durations, scheduling under uncertainty, sample average approximation, constraint programming, mixed-integer linear programming}



\maketitle

\section{Introduction} \label{sec:introduction}
    
    \begin{figure}[b!]
        \centering
        
        \includegraphics[width=0.5\linewidth]{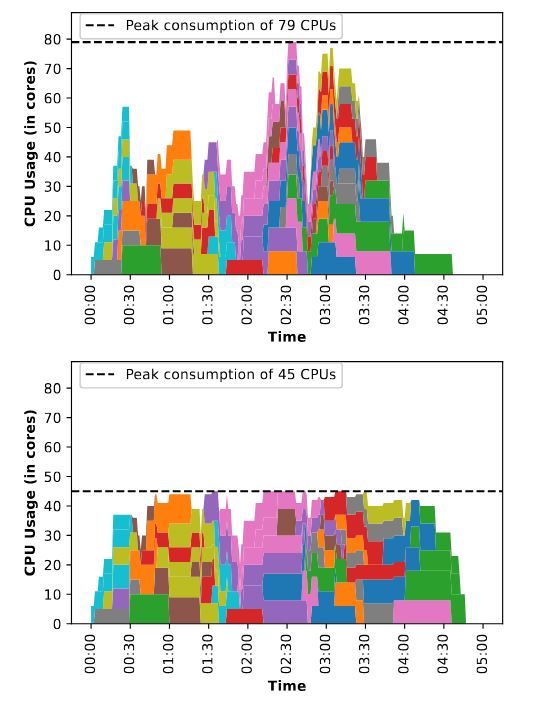}\hfill
        \caption{An example showing 50 jobs before and after computing a schedule minimizing the CPU cores usage. Each job is represented by a different color and has a random duration selected between 10 minutes and 50 minutes. The CPU cores usage is selected randomly between 5 and 10 CPU cores per job.
         }
        
        \label{fig:schedule}
    \end{figure}
    

    In grid-compute environments, where thousands of jobs are executed daily, it is important to ensure that jobs meet their requested deadlines and that computing resources, such as CPU and GPU servers, are used efficiently. 
     Jobs are run on a daily, weekly, or monthly basis as they perform a variety of tasks for traders, portfolio managers, and risk analysts working in our organization. Daily jobs do market data retrieval, risk evaluation, and portfolio management calculations for a variety of securities, such as  stocks, options, and futures. The duration and resource usage of a job vary due to inherently time-varying input parameters, for example, a busy trading day because of stock earnings announcements would lead to a high trading activity and generate a relatively larger amount of market data. This would increase the duration and CPU usage of the jobs compared to an ordinary day.

   A critical first step is determining the amount of resources to allocate for a set of jobs, a process known as capacity planning. The goal of capacity planning is to provide a reliable estimate of the necessary resources to ensure timely job completion while minimizing resource usage. Overestimation of CPU needs can lead to underutilized resources, resulting in unnecessary costs, whereas underestimation may cause delays in job completion, adversely impacting critical stakeholders. The advantage of doing capacity planning and scheduling together is that efficient scheduling enables better resource utilization, reducing capacity requirements while also ensuring timely job completion.

Our organization's cost model is designed in a way that the cost associated with executing jobs is directly proportional to the peak number of CPU cores utilized at any point during the day. Therefore, the objective of our capacity planning and scheduling problem is to minimize the peak CPU cores required for executing a set of jobs on a given day, ensuring both cost efficiency and timely job completion. Thus, our problem differs from the traditional objective of job/project scheduling problems that minimize the makespan (i.e., the interval between the earliest job start time and the latest job end time).

    
    Our stakeholders, who use the grid compute environment and define the jobs, are typically concerned about the completion time of the jobs and not necessarily when they start. Before our proposed solution, the start time of the jobs were typically chosen manually in a conservative manner and primarily driven by convenience, for example, they coincided with the start of office hours on a particular day. This led to inefficient usage of our on-prem servers and motivated us to optimize the capacity planning and scheduling process. Figure \ref{fig:schedule} shows an instance of the impact an intelligent scheduler can have. Say, some jobs with a typical runtime duration of 1 hour and with a deadline of 5 am are scheduled manually to start at 2 am. This leads to the following two issues: (a) if many users follow the same pattern, we observe a high-peak CPU usage in compute utilization for a short duration after 2 am and very low usage later, (2) jobs with much earlier deadlines can be starved of resources, leading to deadline violations and unsatisfied stakeholders. The stochastic nature of job duration and resource usage makes this situation even worse. Therefore, an intelligent job scheduler plays a key role in distributing the workload while ensuring the compute capacity is used efficiently and the jobs are completed on time.

  While there are scheduling algorithms that incorporate duration uncertainty (see section~\ref{sec:related_work}), scheduling algorithms typically do not handle stochastic CPU usage. Our contribution lies in effectively handling {\em{both}} (a) the uncertainty in the duration of the jobs, and also, (b)  the uncertainty in the resource usage of the jobs, effectively.
    Furthermore, our objective is to minimize peak resource usage, distinguishing us from a majority of existing methods that primarily focus on minimizing makespan. Finally, accurate capacity estimation before job execution is essential for us so that the appropriate amount of resources are assigned. In our grid compute infrastructure, resources may not be readily available on demand during execution, making reliable forecasting of capacity requirements crucial for delivering high-quality service.

   
    In summary, our contributions are as follows:
    
    \begin{itemize}
        \item Three approximate approaches that provide capacity estimation {\em and} generate start-time job schedules. They do so while taking into consideration the stochastic behavior in the duration and CPU usage of jobs with temporal deadline constraints:
        \begin{itemize}
            \item a deterministic estimator-based Constraint Programming approach,
            \item a deterministic estimator based Mixed Integer Linear Programming (MILP) approach 
            \item COSPiS (\underline{C}apacity \underline{O}ptimization and \underline{S}cheduling with \underline{P}a\underline{i}r \underline{S}AA): a Sample Average Approximation (SAA)-based approach that builds a constraint-programming model by doing pair sampling, and,
            \citep{chen2020robust}.
        \end{itemize}
         
        \item We conducted experiments with real jobs and evaluated all approaches using four metrics: peak reduction, capacity under-estimation error, capacity over-estimation error, and degree of deadline violation. 
        \item Our proposed COSPiS shows the best performance and achieves up to 41.6\% estimated peak reduction with very low capacity estimation errors on our organization's real job data. 
        \item A proposed end-to-end job execution pipeline integrated with our approaches for capacity planning and job scheduling in our large multinational organization.
    \end{itemize}
    
    The paper is organized as follows. We start by discussing the related work in section ~\ref{sec:related_work}. Then, we define the capacity planning and job scheduling problem with uncertainty in section~\ref{sec:problem_description}. We describe our proposed approaches in section~\ref{sec:our_approach} and present our experimental evaluation in section~\ref{sec:experimental_results}. Finally, we discuss our end-to-end job execution workflow in section~\ref{sec:deployed_app} and conclude in section~\ref{sec:conclusion}. All the notations used throughout the paper are summarized in the Appendix.

\section{Related Work}\label{sec:related_work}

    Our problem of capacity planning and scheduling with uncertainty is related to many well-studied problems in literature. In this section, we refer to these related areas of work and describe how our problem and solution is different. The main related areas are job scheduling with duration uncertainty, resource-constrained project scheduling problem (RCPSP), resource investment problem (RIP), and, optimization in grid-compute frameworks.
    Most of the studies we examined have adapted the deterministic multiprocessor job scheduling problem \citep{morihara1983bin} by incorporating elements of uncertainty in various ways. The multiprocessor job scheduling problem considers multiple types of resources and jobs. Even without uncertainty, this is an NP-HARD problem and the classical bin-packing problem can be reduced to the multiprocessor job scheduling problem, as shown in \citep{morihara1983bin}. We summarize our literature review in Table~\ref{table:related_work}. 

    \begin{table}[h!]

\begin{tabular}{|c|c|c|c|c|c|}
    \hline
    \textbf{Reference} & \textbf{Capacity } & \textbf{Objective} & \textbf{Duration} & \textbf{Resource} & \textbf{Area} \\
     & \textbf{Planning} & & \textbf{Uncertainty} & \textbf{Uncertainty} &  \\
    \hline
    \cite{malewicz2005parallel, azar2021flow, gopalakrishnan2022assignment, bidot2009theoretic, bidot2005general} & & Minimize & \checkmark & & SU \\
     \cite{ radhamani2013performance, ma2016resource} & & Makespan &  & &  \\
    \hline
     \cite{neumann2002project, habibi2018resource, hartmann2022updated, zhou2021stochastic, song2019sampling} & & Minimize & \checkmark & & RCPSP \\
      \cite{rostami2018new, creemers2015minimizing, creemers2016preemptive, varakantham2016proactive} & & Makespan & \checkmark & &  \\
    \hline
    \cite{xiong2013knowledge, hsu2005new, shadrokh2007genetic, bao2022robust, gerhards2020multi} & \checkmark & $min(Peak)$ & \checkmark & & RIP \\
    \hline
    \cite{liu2021online, chen2016uncertainty, yin2022stochastic} & & $min(Peak)$ & \checkmark & & CloudO \\
    \hline
    \cite{tran2018multi, li2023cost} & & $min(Peak)$ &  & & CloudO \\
    \hline
    Ours & \checkmark & $min(Peak)$& \checkmark & \checkmark & RIP \\
    \hline
\end{tabular}
\caption{Summary of our literature survey showing the focus of different papers. SU: Job Scheduling with Uncertainty, RCPSP: Resource-Constraint Project Scheduling Problem, RIP: Resource Investment Problem, CloudO: Cloud Optimization.}
\label{table:related_work}
\end{table}

    \paragraph{Job Scheduling with Duration Uncertainty}
    Several works \cite{malewicz2005parallel, azar2021flow} can handle uncertainty in job duration but {\em not} in the resource usage. They minimize the problem makespan $T$, for e.g., 
    \cite{malewicz2005parallel} presents an approximate solution that does parallel scheduling with complex directed acyclic graphs; \cite{azar2021flow}  can handle an online arrival of new jobs running on a single resource (i.e., one CPU). However, these approaches do not accommodate the job deadline constraints that we have.
         \cite{gopalakrishnan2022assignment} proposes a makespan minimization approach for task scheduling with duration uncertainty where jobs can be interrupted in the middle of execution and be restarted later.
         Some approaches \cite{bidot2009theoretic, bidot2005general} focus on finding globally optimal solutions for stochastic jobs running on a single unit resource (i.e., only one CPU/GPU core). In contrast, our solution accommodates multiple available resources (CPU/GPU cores) to run the jobs, and takes into account the stochastic resource requirements. 

          Some works \cite{radhamani2013performance, ma2016resource, chen2020robust} attempt to use genetic programming-based algorithms for job scheduling with uncertainty. \cite{radhamani2013performance} developed an efficient heterogeneous multi-core scheduling strategy using genetic programming. \cite{ma2016resource} is another such approach. Both of them focus on handling uncertainty only in job duration.  \cite{chen2020robust} presents a distributionally robust optimization algorithm using smart sampling using techniques, such as using $k$ nearest neighbours.

          Finally, linear programming has been used to model scheduling problems.   \cite{oddi2015multi} proposed a multi-objective large neighbourhood search algorithm to efficiently find high quality approximations of the solution Pareto front in multi-objective scheduling problems.

    \textbf{Summarizing} the area of job scheduling with uncertainty, the above approaches do not consider uncertainty in the resource usage of the jobs. Additionally, they do not optimize for peak resource usage, which can limit their effectiveness in capacity planning.
    
    \paragraph{Resource-Constraint Project Scheduling}
    
    We focus on the Resource-Constraint Project Scheduling Problem (RCPSP) \cite{neumann2002project}, due to its relevance to our problem of capacity planning and scheduling under uncertainty. RCPSP focuses on allocating limited resources to project activities over time to minimize the overall project duration while adhering to project constraints and resource limitations. However, RCPSP {\em differs} from our problem in that imposes constraints on the resource usage, and does not consider resource usage as an objective. Instead, RCPSP optimizes the project makespan. 
    
    \cite{neumann2002project, habibi2018resource, hartmann2022updated} provide a thorough review of RCPSP covering various methods for jobs with deterministic resource usage and durations. We discuss the studies that attempt to solve RCPSP with stochastic job durations. \cite{creemers2015minimizing, creemers2016preemptive}  minimize the expected makespan of a project with stochastic activity durations under resource constraints. \cite{rostami2018new} generates a stochastic policy where scheduling times are decided online. 
       
     \cite{zhou2021stochastic} uses a MILP optimization formulation for the stochastic project scheduling problem with time-varying weather conditions. \cite{varakantham2016proactive} presents a Sample Average Approximation (SAA) based approach for minimizing project makespan. \cite{song2019sampling} presents a proactive sampling approach for jobs whose duration uncertainty is dependent on their start times. None of these approaches consider uncertainty in resource usage or minimize peak resource usage.
         
      \textbf{Summarizing} the area of resource-constraint project scheduling, we noticed that the existing algorithms for RCPSP with uncertainty primarily focus on incorporating uncertainty in the durations being optimized, while not considering the uncertainty in resource usage. Our solution, however, takes into account uncertainty in both dimensions: duration and resource usage.
      

    \paragraph{Resource Investment Problem}

    There is a similarity between our problem of capacity planning and the resource investment problem (RIP) \citep{gerhards2020multi, xiong2013knowledge,  hsu2005new, shadrokh2007genetic, bao2022robust} which also aims to minimize the total amount of resources used, and thus able to provide capacity estimations. \cite{xiong2013knowledge} introduces the Stochastic Extended Resource Investment Project Scheduling Problems proposes a knowledge-based multiobjective evolutionary algorithm for solving it, however, it only considers uncertainties in job duration and optimizes for makespan, cost, and robustness.  \cite{bao2022robust} presents a robust optimization model for minimizing resource investment costs in an aircraft assembly line under uncertain processing times, and manages the trade-offs between investment cost, completion time, and uncertainty levels.
    \citep{xiong2013knowledge,  hsu2005new, shadrokh2007genetic, bao2022robust}  assume that resource usage is deterministic with every run consuming a fixed amount of resources. In our setup, the resource requirements are uncertain. 


     \textbf{Summarizing} the area of resource investment problems, we observed that while existing techniques optimize resource usage, they typically do not account for uncertainties in the resource usage of the jobs. Addressing the resource usage uncertainty is a key requirement of our approach.

    \paragraph{Grid Compute Optimization}

    Finally, we also reviewed works \cite{liu2021online, chen2016uncertainty, tran2018multi, li2023cost, yin2022stochastic} in optimizing workflows in large computing infrastructures. \cite{liu2021online} proposes an online multi-workflow scheduling framework, which considers jobs with random arrivals and uncertain task execution times. It provides a way of minimizing rental costs for cloud and reducing deadline violation probability. However, they do not consider the uncertainty in the resource usage of the tasks. \cite{li2023cost}  minimizes costs with job deadline constraints when executing workflow applications in the cloud without any uncertainty. Also, unlike our fixed pricing model (where the capacity needs to be decided beforehand), both \cite{liu2021online} and \cite{li2023cost} are designed for a pay-as-you-go pricing model, and do not need to provide a capacity estimate beforehand.

\cite{chen2016uncertainty} proposes a technique for scheduling real-time workflows in cloud environments with uncertain task execution times. It introduces an uncertainty-aware scheduling framework which exploits both proactive and reactive strategies for reducing costs and improving resource utilization. \cite{yin2022stochastic} develops a stochastic scheduling algorithm for a hybrid cloud architecture, aiming to maximize the private cloud provider's profit and ensure quality service by accounting for the uncertainty of task execution times. While both \cite{chen2016uncertainty} and \cite{yin2022stochastic} do consider duration uncertainty and focus on optimizing costs, they do not do capacity planning because they rely on a pay-as-you-go pricing model. 

\cite{tran2018multi} does resource-aware scheduling where, along with start-time, machine assignment is a part of their scheduling policy. Their three-step algorithm efficiently plans and schedules computational jobs within large-scale data centers, highlighting the benefits of considering job and machine assignment in scheduling decisions. However, they also do not do capacity planning or consider uncertainty.

\textbf{Summarizing,} while much of the research in cloud computing optimization tends to focus on a pay-as-you-use model, where additional resources can easily be allotted on demand, we are concerned with capacity estimation for on-premises grid computing frameworks. In our settings, resources may not be readily available on demand which makes  reliable forecasting of capacity requirements an essential feature for delivering high quality of service.

\section{Problem Description} \label{sec:problem_description}

    In this section, we define a job and our capacity planning and job scheduling problem (COS for \underline{C}apacity \underline{O}ptimization and \underline{S}cheduling) which considers the duration and CPU usage uncertainty for all jobs.

   \subsection{Job.} A job $b$ is defined as $b = (q, f, u, D, J, R)$,
       where,
       \begin{itemize}
           \item $q$: the requested start time of job $b$,
           \item $f$: a measure of flexibility that indicates the maximum amount of time the job $b$ can be delayed to start after its requested start time $q$,
           \item $u$: the latest completion time (deadline) of job $b$,
           \item $D$: a list consisting of the recorded durations or running times of job $b$'s previous executions from historic data,
           \item $J$: the set of jobs that job $b$ depends on consists of all the jobs in $J$. Job $b$ can only start once all the jobs in $J$ have been completed. In this context, the jobs in $J$ are referred to as the parents of job $b$,
                      \item $R$: the history of the number of CPU cores utilized by job $b$.
       \end{itemize}

\begin{figure}[h!]
   \centering
   \includegraphics[width=0.5\columnwidth]{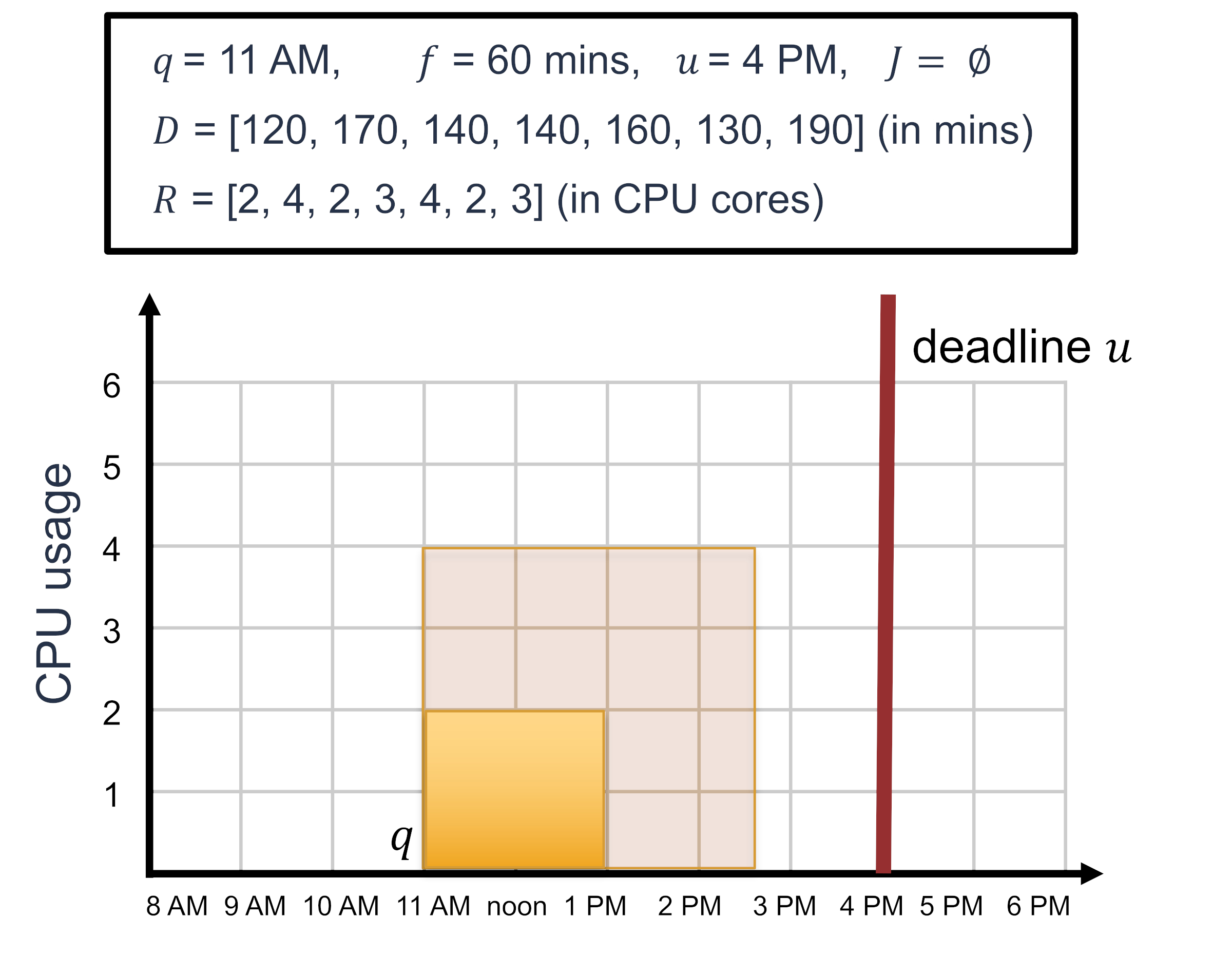}
   \caption{An example of a single job with stochastic duration and CPU usage. The orange square represents the job, and the shaded region represents the uncertainty in the two dimensions.}
   \label{fig:single_job_with_uncertainty}
   \end{figure}

    Figure~\ref{fig:single_job_with_uncertainty} shows an example of a job $b$.
   \begin{figure*}[h!]
   \centering
   \includegraphics[width=0.9\linewidth]{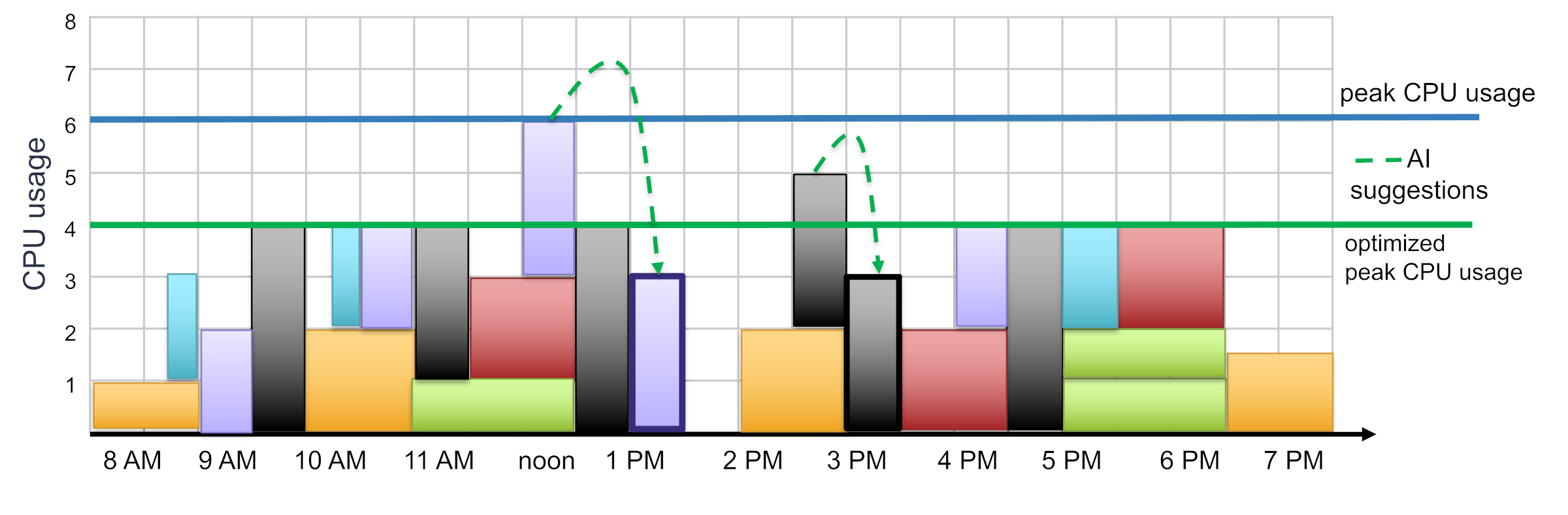}
   \caption{This is an example of our Capacity Planning and Scheduling Problem (COS). Each rectangle in the figure represents a job instance with varying duration and CPU usage, as shown in Figure~\ref{fig:single_job_with_uncertainty}. If there are multiple rectangles of the same color, it means that the same job was submitted multiple times during the day with different contexts and input parameters.}
   \label{fig:scheduling_problem}
   \end{figure*}
   
   \subsection{Capacity Planning and Scheduling Problem with Uncertainty (COS)} Given a set of $n$ jobs $B_n=\{b_j\}_{j=1}^n$, where $b_j = (q_j, f_j, u_j, D_j, J_j, R_j)$, a schedule is defined as $S_n = (s_1, s_2, ..., s_n)$, where $s_j$ is the scheduled start time of job $b_j$. We have two types of uncertainty; (1) the uncertainty associated with the job duration, and, (2) the uncertainty around the resource usage of each job (number of CPU cores used in our case). In other words, the running time and the amount of resources used for a given job would vary during execution. All jobs need to run within a {\em fixed} timespan (makespan) of $T$. The \textbf{Capacity Planning and Scheduling Problem (COS)} with uncertainty is to estimate the peak resource usage, $p$, and, find a start-time schedule $S^*_n = (s^*_1, s^*_2, ..., s^*_n)$ for $B_n$ within a fixed makespan of $T$ that \textit{minimizes the maximum resource usage ($p$) across all jobs at any time while respecting completion deadline of the jobs}.

   Figure~\ref{fig:scheduling_problem} shows an instance of how the peak CPU usage can be reduced by moving some of the jobs to a later time when there are available CPUs.

\section{Our Approaches} \label{sec:our_approach}
    Before describing our approaches in detail, we should highlight the fact that our notion of optimality is a soft one, and we focus on finding approximate solutions for the COS problem in section~\ref{sec:problem_description}. Instead of aiming for absolute perfection, we focus on finding reasonably good solutions given the inherent uncertainties in our COS problem. Planning for worst-case scenarios, where all jobs take their maximum possible duration and CPU usage, can lead to overly cautious schedules with high CPU allocations. This approach may not be practical or efficient, especially when considering that the actual job execution can encounter new and unforeseen values for duration and CPU usage due to the dynamic nature of jobs and the uncertain environment.

Our analysis of historical data has revealed the presence of rare outliers, characterized by spikes in job duration and CPU usage. These outliers are not representative of typical job executions and may distort the optimization process if given too much emphasis. Therefore, we need an approach that can handle such outlier scenarios without being overly sensitive to them.
Considering this factor, we have decided to adopt approximate optimization approaches that allow us to relax the problem constraints to some extent and find solutions that strike a balance between optimality and practicality.

     We propose three approximate approaches to finding a solution to our COS problem.   Our first approach uses a deterministic estimate (e.g. median, mean, quartile, etc) from the available historical data of job durations ($D$) and CPU usage ($R$). Our second approach incorporates the uncertainty into the model by proactively sampling $D$ and $R$ from the historical data. In these two approaches, the COS problem is modeled using Constraint Programming (CP) with the variables:
    
    \begin{itemize}
        \item $\{s_j\}^n_{j=1}$: a set of integer variables where $s_j$ indicates the start time of job $b_j \in B_n$,
        \item $p$: an integer variable indicating the maximum (peak) number of CPU cores used across all jobs at any time $t \in T$. 
    \end{itemize} 

    Our third approach models the problem as a Mixed-Integer Linear Programming (MILP) with a deterministic estimator. For the MILP formulation, we use the RSOME library (Robust Stochastic Optimization Made Easy) \citep{chen2020robust} to solve it.

    \subsection{Constraint Programming with Deterministic Estimator}
    \label{sec:cp_det_est}
        
        In this approach, an estimate of a given job duration and CPU usage is calculated from historic data. Any estimator function $\boldsymbol{\mathit{f^{est}}}$ can be used to approximate their values. The job $b_j$ is now mapped to  
        \[b^{\boldsymbol{\mathit{f^{est}}}}_j = (q_j, f_j, u_j, \boldsymbol{\mathit{f^{est}}}(D_j), J_j, \boldsymbol{\mathit{f^{est}}}(R_j)),\]
        
        \noindent and $B_n \mapsto B^{\boldsymbol{\mathit{f^{est}}}}_n$. To simplify the notations used, we use $\hat{b}_j$ and $\hat{B}_n$ as shorthands for $b^{\boldsymbol{\mathit{f^{est}}}}_j$ and $B^{\boldsymbol{\mathit{f^{est}}}}_n$. So,
        \begin{equation}
            \hat{b}_j = (q_j, f_j, u_j, \hat{d}_j, J_j, \hat{r}_j) \text{ and } \hat{B}_n = \{\hat{b}_j \}_{j=1}^{n},
            \label{eq:estimated_job}
        \end{equation}

        \noindent where $\hat{d}_j = \boldsymbol{\mathit{f^{est}}}(D_j)$ and $\hat{r}_j = \boldsymbol{\mathit{f^{est}}}(R_j)$ are respectively estimations of the duration and CPU usage of job $b_j$.
        
        \paragraph{(1) Temporal Deadline Constraints} Each job $b_j$ has a completion deadline that needs to be respected.
        
        \begin{equation}
            s_j + {\hat{d}_j} \leq u_j, \quad \forall \hat{b}_j \in \hat{B_n}
        \end{equation}
        
        \paragraph{(2) Job Dependency Constraints} Each job $b_j$ has a set of parent jobs $J_j$ that need to be completed before $b_j$ can start. With $s_p$ as the start-time of a given parent job $b_p$,
        
        \begin{equation}
            s_p + \hat{d}_p \leq s_j, \quad \forall j \in \{1, .., n\}, \forall b_p \in J_j.
        \end{equation}
    
        
        \paragraph{(3) CPU Usage} The maximum number of CPU cores used is formulated in terms of the running time intervals of the jobs, and the cumulative number of CPU cores used at each time point. $x_j = [s_j, s_j + \hat{{d}}_j]$ represents the time interval in which job $b_j$ is running. In constraint programming,  cumulative constraint \cite{cumulative} is represented as $\mathtt{cumulative}(\mathbf{X}, \mathbf{Y}, p)$, where $\mathbf{X} = \{x_j\}_{j=1}^n$ is the set of job runtime intervals, and $\mathbf{Y} = \{\hat{{r}}_j\}_{j=1}^n$ is the set of resource usages for $n$ jobs. This 
        constraint ensures that at each time point $t \in T$, the cumulated resource usage of all jobs currently in progress is upper bound by the value assigned to the variable $p$, i.e.,
        
        \begin{equation}
        \mathtt{cumulative}(\mathbf{X}, \mathbf{Y}, p)  \Leftrightarrow \\ 
            \sum_{\substack{j\in \{1,..., n\},\\ t\in x_j }} \hat{{r}}_j \leq p  \forall t \in T
        \end{equation}
        
        \paragraph{(4) Objective function} The objective is to minimize the maximum resource usage, $p$.  

            \begin{figure*}
    \centering
       \boxed{
        \def\arraystretch{1.5}
            \begin{array}{l}
             \underline{\textbf{Constraint Programming (CP) with Deterministic Estimator}}\\
             
                \text{Job: }b_j = (q_j, f_j, u_j, D_j, J_j, R_j); 
                \text {Set of $n$ jobs, } B_n=\{b_j\}_{j=1}^n\\
                
                \text{Start-time Schedule: } ({\color{red} s_1}, {\color{red} s_2}, ..., {\color{red} s_n})\\
              
                \text{Job Flexibility Constraints: }  {\color{red} s_j} \in [q_j, min\{q_j + f_j, u_j\}]  \\ 
                
                \hline

                \hat{b}_j = (q_j, f_j, u_j, \hat{d}_j, J_j, \hat{r}_j); \hat{B}_n = \{\hat{b}_j \}_{j=1}^{n} \\
                
                {\color{red} s_j} + {\hat{d}_j} \leq u_j \quad \forall \hat{b}_j \in \hat{B}_n; \quad\quad \quad
                 
               {\color{red} s_p} + \hat{d}_p \leq {\color{red} s_j} \quad \forall \hat{b}_p \in J_j \hspace{1 em} \forall \hat{b}_j \in \hat{B}_n\\
                
                \text{Intervals: } {\color{red} x_j} = [{\color{red} s_j},{\color{red} s_j} + \hat{{d}}_j]; \quad \quad
                
                \mathbf{{\color{red} X}} = \{{\color{red} x_j}\}_{j=1}^n,  \mathbf{Y} = \{\hat{{r}}_j\}_{j=1}^n\\

                \mathtt{cumulative}(\mathbf{{\color{red} X}}, \mathbf{Y}, {\color{red} p})\\ 
                  
                \textbf{Objective: }\text{minimize ${\color{red} p}$}\\ 
                
            \end{array}
        }
        \caption{Our constraint programming approach with a deterministic estimator. 
        All variables are highlighted in red. $\mathcal{M}$ is a large positive constant.}
        \label{fig:model_Det}
        \end{figure*}

    \subsection{COSPiS: Constraint Programming with Pair Sample Average Approximation (SAA)}
        
        Our solution using Sample Average Approximation (SAA) technique is based on a proactive sampling strategy inspired by the SORU algorithm  \cite{varakantham2016proactive}. We will highlight the differences between our model and SORU at the end of this subsection. 
        The general idea of the SAA approach is to take a certain number of samples from the history of previous runs of each job (in this scenario, we take samples for both the job duration and the number of CPU cores used), and ensure that the required constraints are satisfied for a relevant sub-set of the samples. Since we are sampling a pair (of duration and CPU usage), we call it pair sampling.
        
        \paragraph{(1) Proactive Sampling} For every job $b_j$, we choose $K$ pair samples from the job's duration and resource usage distribution. After sampling, a $K$ pair sampling of $b_j$ is be presented as:
         
        \begin{equation}
            b^K_j = (q_j, f_j, u_j, {D}^K_j, J_j, {R}^K_j)
        \end{equation}
        
        where ${D}^K_j = \{d_{j1}, ...,d_{jk}\}$ is a set of $K$ samples chosen from job $b_j$'s duration in the past, and  ${R}^K_j = \{r_{j1}, ...,r_{jk}\}$ is the same for resource usage. We extend the same notation to map the set of jobs $B_n$ to $B^K_n$, where ${B}^K_n$ represents the scheduling problem of $n$ jobs, where the duration+resource usage of each job is represented with a set of $K$ samples.

        A small subset of the $K$ samples of $B^K_n$ are allowed to violate the constraints depending on the value of a tolerance parameter, $\alpha$. 
        For example, $\alpha = 0.1$ with 10 samples will mean it is acceptable if 1 out of the 10 samples is not completed within the requested deadline.
        $\alpha=0$ means that we want the start time schedule to meet the deadlines for all samples. 
        For each sample $k$, we define a boolean violation variable,
        \[v_k =
        \begin{cases}
            1 \text{ if the sample $k$ of ${B}^K_n$ is ignored for}\\
            \text{constraints,}\\
            0 \text{ if the sample $k$ of ${B}^K_n$ is respected in} \\
            \text{constraints,}\\
        \end{cases}\]

        Setting the tolerance to zero is equivalent to planning for a situation where all jobs take the maximum possible duration and CPU usage to complete. Focusing on this worst-case scenario will lead to extremely cautious schedules with high CPU allocations. Further, if some samples have a job duration beyond the time deadline of that job, this can lead to infeasible models. The tolerance parameter of the COSPiS approach can be tuned so that this situation is avoided. Our analysis of the historic data shows the presence of high-valued outliers (spikes in job duration and CPU usage) which occur rarely. The SAA approach allows us to have a relaxed problem that is not so sensitive about respecting all such outlier scenarios. Further, even if we plan for the worst-case scenario, due to the dynamic nature of the jobs and uncertainty in the environment, we can encounter new higher values of duration and CPU usage during job execution in real time. This is why we chose to go with a soft optimization approach to handle the uncertainty.

        \paragraph{(2) Temporal Deadline Constraints} We allow a portion of the samples to violate the temporal deadline $u_j$ depending on the value chosen for the variable $v_k$.
        
        \begin{equation}
            s_j + d_{jk} \leq u_j + \mathcal{M} v_k, \quad \forall k \in \{1,...,K\}
        \end{equation}
        
         where $\mathcal{M} \geq T$ is a large positive constant, and $d_{jk}$ is the $k$th item of sample set $D^K_j$.

         Note that a delay in one job can cause a ripple effect and result in delays for other jobs which are dependent on this job. However, the next day when we compute the schedule, this new duration now becomes a part of the historic data and model formulation (part of $D_i$). If we want to update the schedule online during the day, we would remove the constraints related to the start time variables of the jobs which have finished executing and reschedule online.

         \paragraph{(3) Dependency Constraints} Each job $b_j$ has a set of parent jobs $J_j$ that should be completed before $b_j$ can begin. We allow a portion of the samples to violate this requirement depending on the value chosen for the variable $v_k$. 
         
        \begin{equation}
            s_p + d_{pk} \leq s_j + \mathcal{M} v_k,  \quad  \forall b_p \in J_j, k \in \{1,...,K\}
        \end{equation}
        
        We ensure only a limited portion of the samples are violated by having an upper bound on the sum of violation variables:
        \begin{equation}
            \sum_{k=1}^{K}v_k \leq K \alpha
        \end{equation}
        
        where $\alpha$ is a tunable hyperparameter and $K$ is the number of samples taken for each job. Note that we allow job dependency violation only during model formulation and not when jobs are executed on our compute infrastructure following a schedule. During execution, a job cannot start before its parent jobs have been completed. 
        
        \paragraph{(4) CPU Usage} For each sample $k$, we want to compute the maximum resource usage. Similar to the deterministic estimator formulation, we define a list of time-interval variables for all jobs for all samples. The duration of the running time interval of a job in a sample is the duration of the job in the specific sample. 
        $x_{jk} = [s_j, s_j + d_{jk}]$ is the run time interval of job $b_j$ in sample $k$. 
        
        Now, we compute the peak resource usage of each sample using the $\mathtt{cumulative}$ constraint. Let $\mathbf{X}_k = \{x_{jk}\}_{j=1}^{n}$ be the set of the running time intervals of all jobs, and  $\mathbf{Y}_k = \{{R}^K_j\}_{j=1}^{n}$ be the set of resource usage of all jobs in sample $k$. For sample $k$, $\mathtt{cumulative}(\mathbf{X}_k, \mathbf{Y}_k, p_k)$ implies, 
         
        \begin{equation}
              \sum_{j\in \{1,..., n\}, t\in x_{jk} } r_{jk} \leq p_k, \quad \forall t \in T
        \end{equation}
        
        where $p_k$ is peak usage of sample $k$, and $r_{jk}$ is the $k$th item of sample set $R^K_j$. For all samples, the peak usage variable $p_k$ is the upper bound on the cumulated CPU usage at any time $t \in T$, i.e. for any sample $k$, $\mathtt{cumulative}(\mathbf{X}_k, \mathbf{Y}_k, p_k)$.
        
        \paragraph{(5) Objective function} Our objective is to minimize $max\{p_k\}_{k=1}^{K}$.

    \begin{figure*}
    \centering
       \boxed{
        \def\arraystretch{1.5}
            \begin{array}{l}
             \underline{\textbf{COSPiS: CP with Pair Sample Average Approximation (SAA)}}\\
            
                \text{Job: }b_j = (q_j, f_j, u_j, D_j, J_j, R_j); 
                \text {Set of $n$ jobs, } B_n=\{b_j\}_{j=1}^n\\
                
                \text{Start-time Schedule: } ({\color{red} s_1}, {\color{red} s_2}, ..., {\color{red} s_n})\\
              
                \text{Job Flexibility Constraints: }  {\color{red} s_j} \in [q_j, min\{q_j + f_j, u_j\}]  \\ 
                
                \hline
                

                b^K_j = (q_j, f_j, u_j, {D}^K_j, J_j, {R}^K_j); {B}^K_n = \{{b^{K}_j} \}_{j=1}^n \\
                
                {\color{red} s_j} + d_{jk} \leq u_j + \mathcal{M} {\color{red} v_k}\quad \forall b_j \in B^{K}_n\\
                
                 {\color{red} s_p} + d_{pk} \leq {\color{red} s_j} + \mathcal{M} {\color{red} v_k} \quad \forall b_p \in J_j  \forall \hat{b}_j \in \hat{B}_n \forall k \in [1, K] \\
                
                \text{Intervals: } {\color{red} x_{jk}} = [{\color{red} s_j}, {\color{red} s_j} + d_{jk}]; \quad \quad
                
                {\color{red} \mathbf{X}_k} = \{{\color{red} x_{jk}}\}_{j=1}^{n}, \mathbf{Y}_k = \{{R}^K_j\}_{j=1}^{n} \quad \forall k \in [1, K]\\
                
                \mathtt{cumulative}({\color{red}\mathbf{X}_k}, \mathbf{Y}_k, {\color{red} p_k}) \quad \forall k \in [1, K] \\

                \text{Boolean tolerance variables: }\sum_{k=1}^{K}{\color{red} v_k} \leq K \alpha\\
                
                \textbf{Objective: }\text{minimize }max\{{\color{red} p_k}\}_{k=1}^{K}\\ 

            \end{array}
        }
        \caption{COSPiS: Our proposed approach using constraint programming and SAA (Sample Average Approximation) 
        All variables are highlighted in red. $\mathcal{M}$ is a large positive constant.}
        \label{fig:model_COSPiS}
        \end{figure*}

    Our approach is different from SORU \citep{varakantham2016proactive} in the following ways:
        
        \begin{itemize}
            \item we consider both the uncertainty in the number of CPU cores used and the duration while SORU only considers the latter. Furthermore, our technique can be generalized to any number of variables with uncertainty in a scheduling problem.
            \item we minimize the maximum number of CPU cores used whereas SORU minimizes the makespan of the schedule.
            \item we have the job flexibility and dependency constraints, adding more complexity to our problem.
           \item SORU uses a Mixed Integer Linear Programming (MILP) model while we use Constraint Programming. 
        \end{itemize}
       
The approach we present in the next section is MILP-based using similar auxiliary variables as SORU. However, with preliminary experiments, we found constraint programming to converge faster to feasible schedules than MILP with respect to the number of jobs. As a result, we did not go forward with implementing an SAA version of the MILP model.

    

    \subsection{Mixed Integer Linear Programming (MILP) with  Deterministic Estimator}
    
     To formulate the MILP model, we use the same deterministic mapping for the jobs, $B_n \mapsto \hat{B}_n$ as defined in Equation~\ref{eq:estimated_job} using the estimator function $\boldsymbol{\mathit{f^{est}}}$. The temporal deadline constraints, dependency constraints, and objective function are also the same as Section~\ref{sec:cp_det_est}. 
     
     \noindent \textbf{CPU Usage.} To calculate the maximum CPU usage at any time, we observe that the CPU usage can increase only when a new job begins. We use two sets of Boolean variables $\delta^1_{ji}$ and $\delta^2_{ji}$ to determine if $b_i$ is executing when $b_j$ starts.  
    \begin{equation}
         {\delta^1_{ji}} \geq ({s_j} - {s_i} + 1) / \mathcal{M} \quad \forall {b}_j, {b}_i \in {B}_n,
    \end{equation}
    \begin{equation}
                {\delta^2_{ji}} \geq ({s_i} + \hat{d}_i - {s_j}) / \mathcal{M}\quad \forall {b}_j, {b}_i \in {B}_n,
    \end{equation}
where, $\mathcal{M}$ is a large positive constant. If $\hat{b}_j$ starts after $\hat{b}_i$ has completed, then feasible values are $\delta^1_{ji} = 1$, and $\delta^2_{ji} = 0$ or 1. If $\hat{b}_j$ starts during execution of $\hat{b}_i$, then $\delta^1_{ji} = \delta^2_{ji} = 1$. If $\hat{b}_j$ starts before $\hat{b}_i$ starts, then $\delta^1_{ji} = 0$ or 1, and $\delta^2_{ji} = 1$. 
Now, we will use $\delta^1_{ji}$ and  $\delta^2_{ji}$ to compute the contribution to total CPU usage (denoted using $res_{ji}$) of $\hat{b}_i$ when $\hat{b}_j$ starts.
  \begin{equation}
                {res_{ji}} \leq \delta^1_{ji}\hat{r}_i \text{ and } {res_{ji}} \leq \delta^2_{ji}\hat{r}_i \quad \forall \hat{b}_j, \hat{b}_i \in \hat{B}_n
    \end{equation}
     \begin{equation}
                {res_{ji}} \geq  \hat{r}_i - (2 - \delta^1_{ji} - \delta^2_{ji})\mathcal{M}\quad \forall \hat{b}_j, \hat{b}_i \in \hat{B}_n
    \end{equation}
    Then, the peak CPU usage $p$ has a lower bound which is the total CPU usage when $\hat{b}_j$ starts:
     \begin{equation}            
                \hat{r}_j + \sum_{i \neq j}res_{ji} \leq { p} \quad \forall\hat{b}_j, \hat{b}_i \in \hat{B}_n
    \end{equation}

\paragraph {Job flexibility Constraints} Finally, we are restricted to moving a job only up to a fixed amount defined by its flexibility $f$. This is because rescheduling a job from its requested start time $q$ needs approval from the user who submitted it. So, the start time,
        \begin{equation}
             s_j \in [q_j, min\{q_j + f_j, u_j\}] \quad \forall b_j \in B_n
        \end{equation}

         \begin{figure*}
    \centering
       \boxed{
        \def\arraystretch{1.5}
            \begin{array}{l}
             \underline{\textbf{Mixed Integer Linear Programming 
                (MILP) with Deterministic Estimator}}\\

                \text{Job: }b_j = (q_j, f_j, u_j, D_j, J_j, R_j); 
                \text {Set of $n$ jobs, } B_n=\{b_j\}_{j=1}^n\\
                
                \text{Start-time Schedule: } ({\color{red} s_1}, {\color{red} s_2}, ..., {\color{red} s_n})\\
              
                \text{Job Flexibility Constraints: }  {\color{red} s_j} \in [q_j, min\{q_j + f_j, u_j\}]  \\ 
                
                \hline
                
               
                 \hat{b}_j = (q_j, f_j, u_j, \hat{d}_j, J_j, \hat{r}_j); \hat{B}_n = \{\hat{b}_j \}_{j=1}^{n} \\
                
                {\color{red} s_j} + {\hat{d}_j} \leq u_j \quad \forall \hat{b}_j \in \hat{B}_n\\
                 
               {\color{red} s_p} + \hat{d}_p \leq {\color{red} s_j} \quad \forall \hat{b}_p \in J_j \hspace{1 em} \forall \hat{b}_j \in \hat{B}_n\\
                
             \forall \hat{b}_j, \hat{b}_i \in \hat{B}_n:\\
             
               {\quad\quad \color{red}\delta^1_{ji}} \geq ({\color{red}s_j} - {\color{red}s_i} + 1) / \mathcal{M}; \quad \quad
               
                {\quad\quad\color{red}\delta^2_{ji}} \geq ({\color{red}s_i} + \hat{d}_i - {\color{red}s_j}) / \mathcal{M} \\

                {\quad\quad\color{red} res_{ji}} \leq {\color{red}\delta^1_{ji}}\hat{r_i}; \quad \quad \quad \quad \quad \quad
                  
                {\quad\quad\color{red} res_{ji}} \leq {\color{red}\delta^2_{ji}}\hat{r_i} \\
                
                {\quad\quad\color{red} res_{ji}} \geq  \hat{r_i} - (2 - {\color{red}\delta^1_{ji}} - {\color{red}\delta^2_{ji}})\mathcal{M} \\
                
                \quad\quad \hat{r_j} + \sum_{i \neq j}{\color{red}res_{ji}} \leq {\color{red} p}  \\
                
                \textbf{Objective: }\text{minimize ${\color{red} p}$}\\

            \end{array}
        }
        \caption{Our MILP formulation with  deterministic estimator. 
        All variables are highlighted in red. $\mathcal{M}$ is a large positive constant.}
        \label{fig:model_MILP}
        \end{figure*}

Our MILP formulation with a deterministic estimator has more variables and constraints than the CP models and takes longer to converge as we will see in Section~\ref{sec:experimental_results}. Although it can be extended with SAA, we observed that MILP with a median estimator takes much longer (5-10 times) to converge than our CP approaches. So, we did not develop an SAA model for MILP as it would add more variables and constraints to the model, and the long duration hinders its use for systems in production.

\section{Experimental Evaluation} 
\label{sec:experimental_results}

  In this section, we evaluate the performance of the approaches described in section \ref{sec:our_approach} on a real dataset of COS (Capacity Planning and Scheduling) problems from our organization. We start by describing our evaluation metrics, our dataset and experimental setup, followed by the presentation and analysis of our experimental results.
  
 \subsection{Evaluation Metrics}
 \label{sec:performance_metrics}
    
        In order to measure the solution quality and capture how reliable our capacity estimations are (i.e., how well we handle uncertainty), we define the following four metrics: (1) Peak Reduction, (2) Capacity Under-Estimation Error, (3) Capacity Over-Estimation Error, and (3) Degree of Deadline Violation. These metrics capture the cost reduction and quality of service (QoS) that we deliver to the end users. By optimizing these metrics, we align with business goals, of reducing costs and completing jobs by their deadlines. Separating capacity under-estimation and over-estimation errors is crucial for us because they have distinct impacts: under-estimation can lead to CPU shortages and operational delays for stakeholders, while over-estimation can result in unnecessary costs and inefficiencies. By examining each error type individually, business can balance cost-efficiency and quality of service.             
        
        \paragraph{Peak Reduction} After an optimization approach is used to compute a schedule for a set of jobs, the jobs are executed following the schedule on our grid-compute environment. Due to the stochastic nature of the jobs, the peak CPU usage observed in real-time is most likely to be different than the value predicted by an approach. The peak reduction measures the amount by which the observed peak by an optimization model is lower than manual scheduling. Since the day's billable cost is proportional to the peak CPU usage on that day, the peak reduction metric is an indicator of financial cost savings. In terms of our solution’s usefulness and cost-saving indicators, the amount of money saved is directly proportional to the average peak reduction plus close to five manual hours weekly per 200 jobs. As different employees have different wages (inaccessible to us), it is difficult to accurately monetize the manual efforts saved.
        
        \paragraph{Capacity Under-Estimation Error}  The capacity under-estimation error measures how much the observed peak core usage ($p_{real}$) of a schedule exceeds the predicted capacity value ($p_{est}$) by an optimization model. It is calculated by $max\{0, (p_{real} - p_{est})/p_{est}\}$. As an example, consider a scenario for a schedule $S$ which has a predicted peak of 6 CPU. However, after executing schedule $S$, the actual peak CPU usage is observed to be 10 CPU. In this case, the capacity under-estimation error is 4/6. Note that, if the observed peak from execution is 3 CPU (lower than the predicted value of 6 CPU), the capacity under-estimation error will be 0.  
        
        Allocating extra CPUs during execution is costly ($X$ times the cost of pre-allocating, $X\gg1$), and this cost varies throughout the day. In some cases, additional CPUs may be unavailable. Take two schedules, $S1$, and $S2$. $S1$ has estimated peak of 50, capacity under-estimation error = 10/50, and $Cost(S1)=O(50+10*X)$. $S2$ has an estimated peak of 60 and a capacity under-estimation error of zero. Then, $Cost(S2)=O(60+0*X)$.

        With manual scheduling, there is no optimization in place. To deliver high QoS, our organization has stayed on the safe side and allocated a sufficient for every partition based on a leasing agreement with our vendors to deliver high QoS to end-users. One of the motivations of our work was to find how much we can save (peak reduction) without causing deadline violations. For manual scheduling, the capacity under-estimation error is zero but this incurs the problem of CPUs not being utilized effectively as we continue to pay for them. 

            

        \paragraph{Capacity Over-Estimation Error}
        
        The capacity over-estimation error measures how much the estimated peak usage ($p_{est}$) by an optimization model exceeds the observed peak value ($p_{real}$) of a schedule. It is calculated by $max{(0, (p_{est} - p_{real})/p_{est})}$. This metric is critical in evaluating the efficiency of scheduling algorithms, as overestimating peak resource usage can lead to unnecessary resource allocation, resulting in inefficiencies and increased operational costs.  As an example, consider a scenario for a schedule $S$ which has a predicted peak of 10 CPU. However, after executing schedule $S$, the actual peak CPU usage is observed to be 6 CPU. In this case, the capacity over-estimation error is 4/10 CPU. Note that, if the observed peak from execution is 12 CPU (higher than the predicted value of 10 CPU), the capacity over-estimation error will be 0.

Overestimating CPU requirements leads to resource inefficiency, as extra CPUs remain idle, resulting in wasted capacity and increased costs. Allocating excess CPUs incurs unnecessary expenses, as resources are reserved but not utilized. Consider two schedules, $S1$ and $S2$. $S1$ has an estimated peak of 60, a capacity over-estimation error of 10, and $Cost(S1) = O(60)$. $S2$ has an estimated peak of 50 with a capacity over-estimation error of zero. Then, $Cost(S2) = O(50)$.

With manual scheduling, no optimization is in place, and to ensure high QoS, organizations often allocate more resources than needed, leading to over-estimations. This practice guarantees performance but results in many CPUs sitting idle while still incurring costs. One of the motivations of our work was to find how much we can save (average peak reduction) without causing deadline violations. For manual scheduling, the capacity over-estimation error can be high, highlighting the inefficiency and costliness of over-provisioning.
        
        \paragraph{Degree of Deadline Violation} When a job is executed following a particular schedule $S$, it may or may not finish by its requested deadline. The degree of deadline violation metric measures the amount of delay of each job completion ($u_{real}$) with respect to the job's deadline $u$. Similar to the previous metric, this error also can be calculated by $max\{0, (u_{real} - u)\}$. As an example, if the deadline for a job is at 5 pm, however during the execution it actually finishes at 5:15 pm, the degree of deadline violation is 15 mins. Furthermore, if the job finishes at 4:50 pm (earlier than the deadline), the degree of deadline violation will be 0. In our test cases with manual scheduling, the degree of deadline violation is similar to the three approaches with most jobs finishing within the deadline. 
        
\begin{figure*}[h!]
   \centering
   \includegraphics[width=0.9\linewidth]{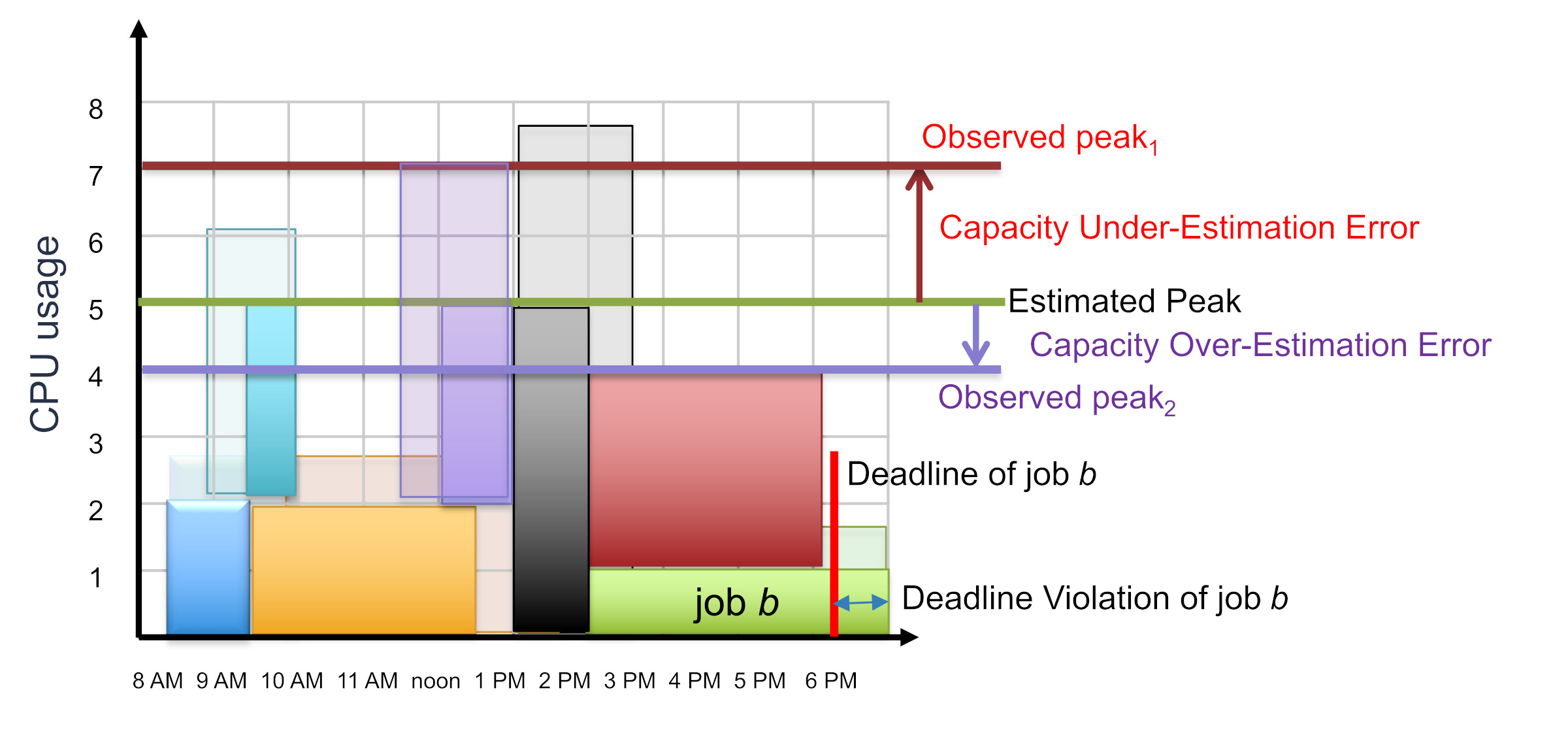}
   \caption{A notional diagram showing three of our evaluation metrics: the capacity under-estimation error, the capacity over-estimation error, and the degree of deadline violation.}
   \label{fig:metrics}
\end{figure*}

        Figure~\ref{fig:metrics} shows a notional diagram of the capacity under-estimation and over-estimation errors, and the degree of deadline violation. For both defined metrics, the lower values indicate better solution quality. 
        
\subsection{DataSet and Experimental Setup}
\paragraph{DataSet}


 In our organization, jobs are scheduled on a daily basis and the execution history of jobs is saved for several months, which we use to construct our dataset of COS problems. In our current dataset, we have 44 problems, with the number of jobs in a problem ranging from 7 to 348, with the median number of jobs in problem = 239. Table~\ref{table:dataset_summary} summarizes the different properties of our dataset.  Since jobs are scheduled on a daily basis, the problem makespan is 1 day. 

 \begin{table}[h!]
\centering
\begin{tabular}{lccccc}
\toprule
\textbf{Property} & \textbf{Min} & \textbf{Max} & \textbf{Median} & \textbf{Mean} & \textbf{Std} \\
\midrule
Number of jobs in a problem & 7 & 348 & 239 & 144 & 122\\
Historic record size of duration and CPU usage & 1 & 3866 & 21 & 98 & 286 \\
Job Durations (secs)     & 21 & 27474 & 152 & 1041 & 2706\\
Uncertainty in Durations (secs)       & 0 & 9411 & 66 & 543 & 1265\\
CPU Usage (cores)          & 3 & 1000 & 23 & 73 & 133 \\
Uncertainty in CPU Usage (cores)       & 0 & 444 & 19 & 56 & 84 \\
Flexibility (secs)        & 2000 & 7258 & 3714 & 4356 & 2250 \\
\bottomrule
\end{tabular}
\caption{Summary of our COS problem dataset of size 44.}
\label{table:dataset_summary}
\end{table}


\paragraph{Experimental Setup}
  We test the three approaches presented in Section~\ref{sec:our_approach}, Det, COSPiS, and MILP, with all the COS problems in our dataset. We compare with another Sample Average Approximate (SAA)-based algorithm, called SORU (from \cite{varakantham2016proactive}). SORU is able to handle uncertainty in job duration (and not in resource usage). However, SORU's objective is to minimize makespan and not the CPU usage. To make a more meaningful comparison, we modified SORU to remove the resource constraints, and optimize for peak resource usage. We call this version of SORU, SORU$^{Pk}$. For approaches that rely on an estimator (Det, MILP, and SORU$^{Pk}$), we experimented with four choices of deterministic estimator, $\boldsymbol{\mathit{f^{est}}}$:
  \[
   \boldsymbol{\mathit{f^{est}}} \in \{ P50(\text{median}), P100(\text{max}), P75(\text{third quartile}), mode \}.
  \]
 Our proposed COSPiS algorithm has two hyperparameters, the number of pair samples to taken from the historic job duration and CPU usage data, and the tolerance, an upper bound on the ratio of samples that may violate the job deadline constraints. We did a hyperparameter study (described in section~\ref{sec:hyperparam_study}) and selected tolerance, $\alpha = 0.4$, and the number of samples, $K = 25$, based on the values of the realized performance metrics.

    For each combination of problem and approach, we did 50 runs to account for the uncertainty in the COS problems.
      We used a CP-SAT solver \citep{ortools} to find the solution to our Constraint Programming models (Det, COSPiS, and SORU$^{Pk}$) and the RSOME \citep{chen2020robust} library for the MILP model. We performed our experiments on a 16 vCPU, 64 GB RAM, and 3 GHz processor. The time limit for the solver for each COS problem was set to 15 mins. 
        
\subsection{Experimental Results} 
        
In this section, we analyze how our proposed capacity planning and scheduling approaches in section~\ref{sec:our_approach} perform with respect to our four performance metrics: peak reduction, capacity under- and over-estimation errors and degree of deadline violation introduced in section~\ref{sec:performance_metrics}.

\paragraph{Peak Reduction}

 \begin{figure}[h!]
         \centering
            \includegraphics[width=0.95\linewidth]{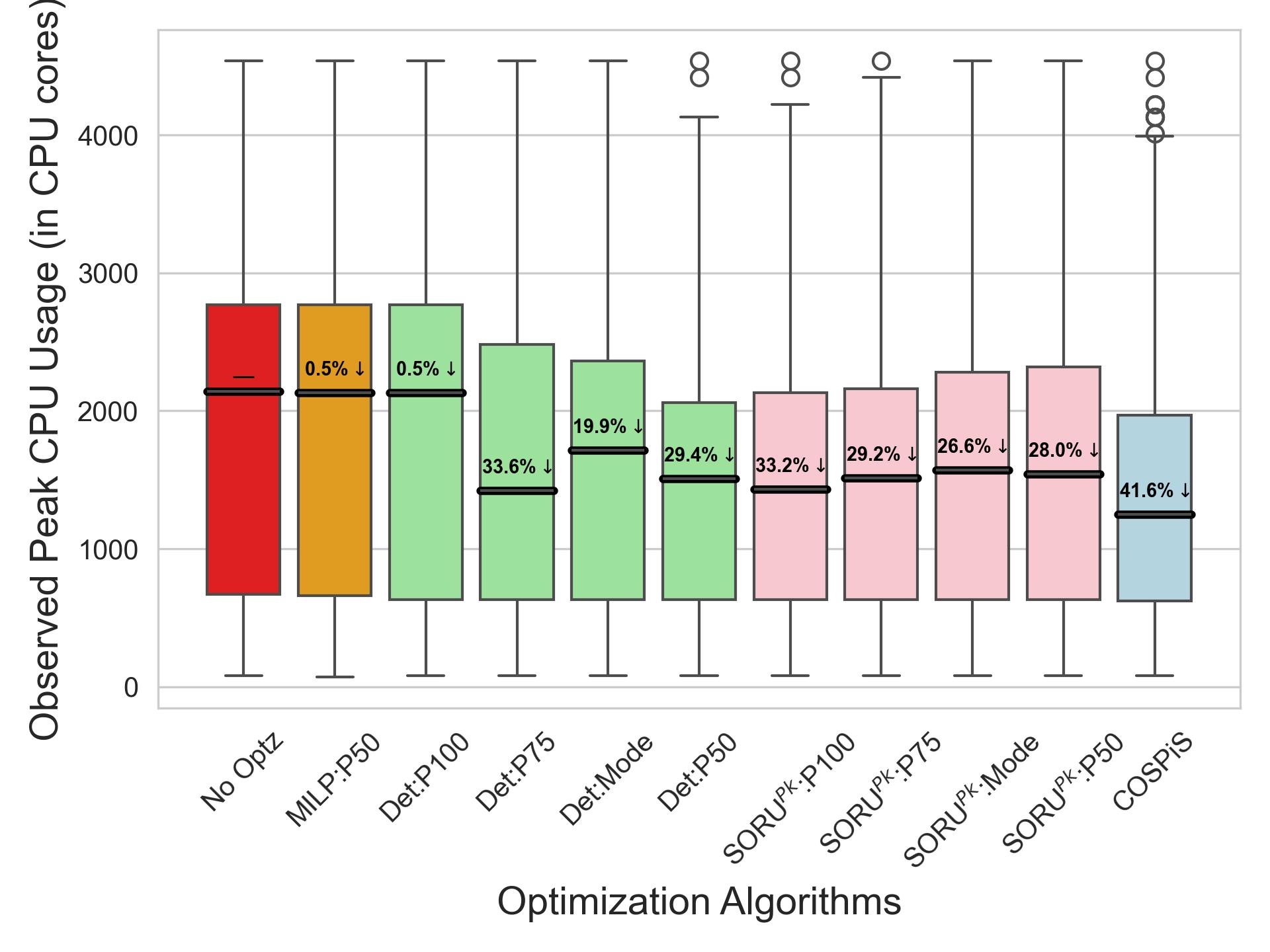}
            \caption{Comparison of observed peak CPU usage across a dataset of 44 COS problems from our organization, with each problem containing between 7 and 348 jobs. The plot shows five categories of approaches: (1) No Optz (manual, with no optimization, shown in red), (2) MILP with a Deterministic Estimator using a median (P50) estimator (orange), (3) Det: Constraint Programming with multiple Deterministic Estimators - max (P100), 75th percentile (P75),  median (P50) and mode (all shown in green), (4) SORU$^{Pk}$, as modified from \cite{varakantham2016proactive} to optimize peak CPU usage, employing estimators for CPU usage - max (P100), 75th percentile (P75),  median (P50) and mode (all shown in pink), and (5) COSPiS (Capacity Planning and Scheduling via Pair Sampling), our proposed algorithm (shown in blue). The percentage (higher  $\rightarrow$ better) label in each box shows the reduction in peak CPU usage compared to the baseline scenario of manual scheduling (no optimization). Our COSPiS approach demonstrates the best performance, achieving a peak reduction of 41.6\%.}
            \label{fig:observed_peak_all_approaches}
\end{figure}

Figure \ref{fig:observed_peak_all_approaches}  shows the efficacy of the different optimization algorithms in reducing peak CPU usage for our dataset of COS problems. 
We consider the manual schedule (no optimization) in our organization as the baseline and measure the peak reduction of the other approaches with respect to it.

The deterministic constraint programming with max estimator (Det:P100) and the MILP approach with a median estimator (MILP:P50) do not show much improvement compared to the no optimization scenario, indicating minimal effectiveness in reducing peak resource demands in these configurations. This is because Det:P100 heavily overestimates the duration and CPU usage, leading to overly cautious capacity planning. The situation is different for the MILP:P50 approach, where even with optimistic (median) estimates it is not able to achieve any improvement in performance. This is because for problem sizes $\geq 50$ (details in section~\ref{sec:detailed_exp_study}, it fails to return a feasible solution within our solver time limit of 15 mins.

For three configurations of the Det approaches (Det:P75, Det:P50, and Det:Mode), we actually see significant reduction in the peak CPU usage. Notably, the 75th percentile (P75) estimator achieves a reduction of 33.6\% with respect to no optimization. This suggests that using slightly risk-averse estimators like P75 can effectively moderate peaks more than P50 and mode estimators in our operational settings with the Det Approach. However, the Det approach cannot handle uncertainty in neither duration nor CPU usage. 

 For the SORU$^{Pk}$ configurations, the trends are different than the Det configurations with the same estimators. This is expected because SORU$^{Pk}$ takes into account the uncertainty in the job duration. Here, the maximum (P100) estimator ranks near the top with a 33.2\% reduction. The lower peak reduction with other estimators (P75 at 29.2\%, P50 at 28\%, and mode at 26.6\%) shows that conservative estimation of the CPU usage works better  SORU$^{Pk}$. However,  SORU$^{Pk}$ does not take into account the uncertainty in the CPU usage like our proposed COSPiS approach.

Markedly, the COSPiS algorithm, our proposed solution, stands out with an impressive 41.6\% reduction in peak CPU usage showcasing its ability to effectively handle the uncertainty in {\em both} duration and CPU usage, which minimizing the peak CPU usage. The algorithm's pair-sampling technique aligns well with the characteristics of our dataset for capacity planning and scheduling.

\paragraph{Capacity Under-Estimation and Over-Estimation Errors}
\begin{figure}[h!]
        \centering
            \centering
            \includegraphics[width=0.65\linewidth]{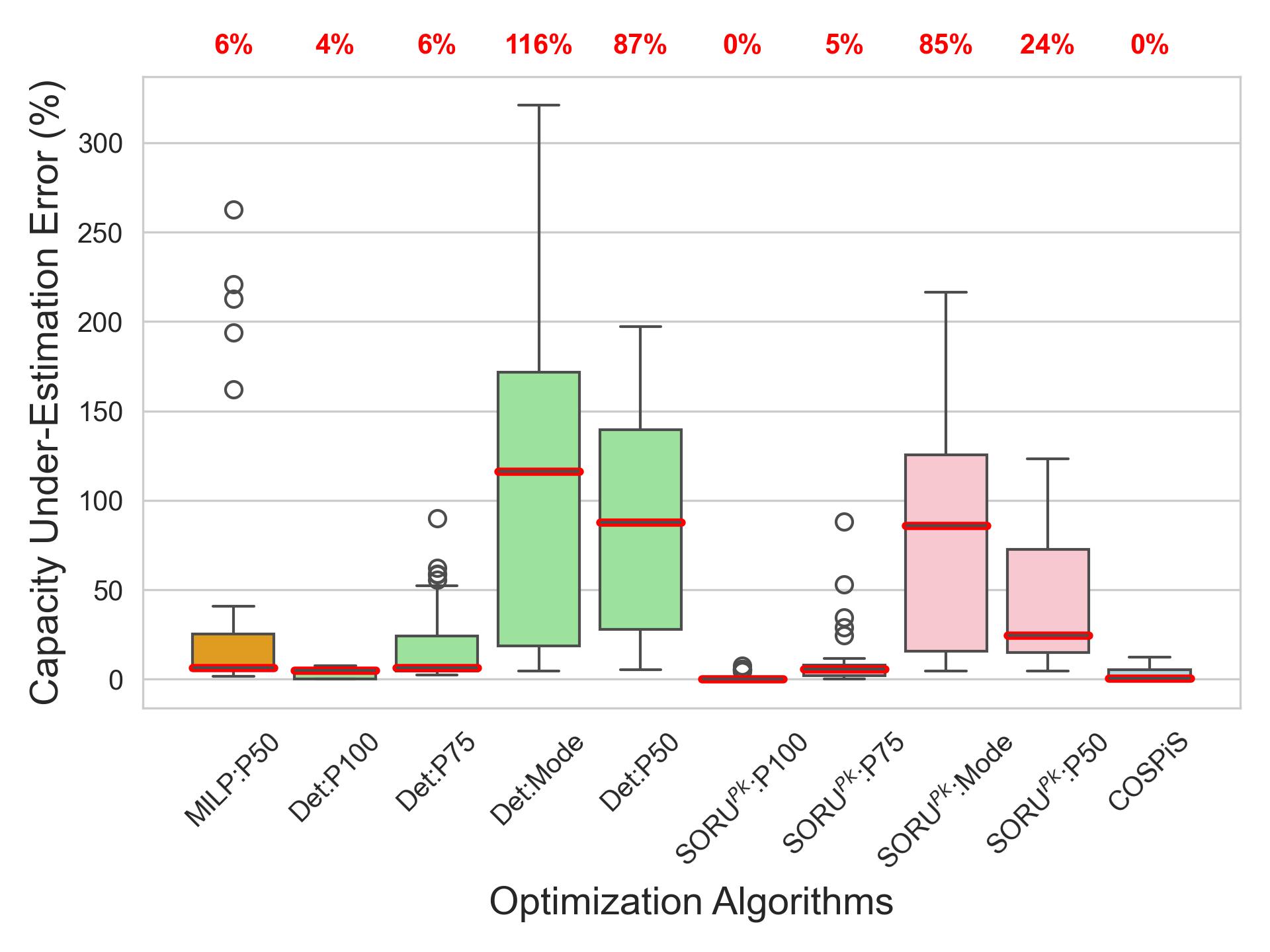}
            \includegraphics[width=0.65\linewidth]{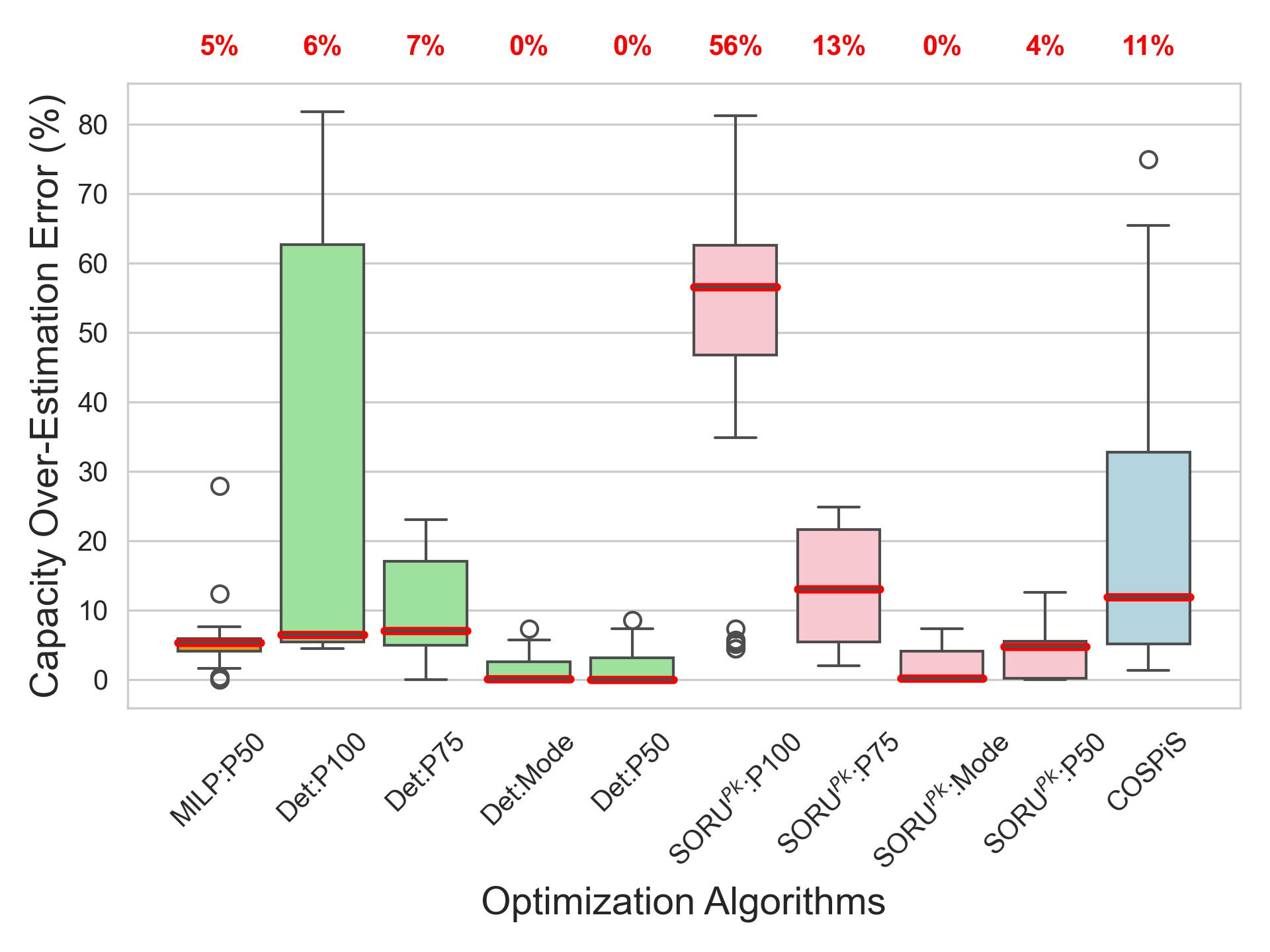}
            \caption{A comparison of the capacity under- (top) and over-estimation (bottom) errors relative to their estimated capacity across a dataset of 44 COS problems from our organization, with each problem containing between 7 and 348 jobs. The plot shows four categories of approaches: (1) MILP with a Deterministic Estimator using a median (P50) estimator (orange), (2) Det: Constraint Programming with multiple Deterministic Estimators - max (P100), 75th percentile (P75),  median (P50) and mode (all shown in green), (3) SORU$^{Pk}$, as modified from \cite{varakantham2016proactive} to optimize peak CPU usage, employing estimators for CPU usage - max (P100), 75th percentile (P75),  median (P50) and mode (all shown in pink), and (4) COSPiS (Capacity Planning and Scheduling via Pair Sampling), our proposed algorithm (shown in blue). The \%tage (higher $\rightarrow$ worse) label on top of each box shows the median error. 
            }
         \label{fig:capacity_estimation_error_all_approaches}
    \end{figure}

Figure~\ref{fig:capacity_estimation_error_all_approaches} shows the capacity under-estimation error and over-estimation errors for the different optimization algorithms in our experimental setup. We measure the errors relative the the estimated peak values by the respective algorithms.

In terms of capacity under-estimation error, Det:P50 and SORU$^{Pk}$:P50 show relatively high error rates at 87\% and 24\% respectively. The mode estimator performs the worst, with Det:Mode exhibiting a considerable underestimation error at 116\%, and SORU$^{Pk}$:Mode with an error of 85\%. The max estimator under both constraint programming (Det:P100) and the SORU$^{Pk}$ method (SORU$^{Pk}$:P100) exhibit minimal capacity under-estimation, with values of 4\% and 0\% respectively. In general, the capacity under-estimation error decreases as the estimators become more conservative. 

When considering capacity over-estimation errors, an opposite pattern (with respect to the estimators) emerges as expected. The mode and median (P50) estimators under both Deterministic and SORU$^{Pk}$ methods tend to have smaller errors, with Det:P50, Det:Mode, and SORU$^{Pk}$:Mode showing no over-estimation error at all, and SORU$^{Pk}$:P50 only with a 4\% error. 

 SORU$^{Pk}$ with max estimator (SORU$^{Pk}$:P100) demonstrates the worst overestimation error of 56\%, which is the highest among all the approaches. This is likely due to the fact that the schedule generated by SORU$^{Pk}$ is better equipped to reduce the peak usage because it considers the job duration uncertainty (unlike the Det approaches) leading to reduced peak at job execution time than estimated.

When looking at the MILP:P50 and COSPiS approaches, we see modest underestimation error rates of 6\% and 0\%, and overestimation errors of 5\% and 11\% respectively. Notably, COSPiS exhibits no underestimation error, and has a higher over-estimation error. This indicates that the COSPiS approach is careful to ensure enough resources are allocated, even maintaining a safer margin. By effectively eliminating the risk of underestimating CPU needs, COSPiS provides a significant benefit to our grid-compute framework ensuring that they are not put in a position to look for additional capacity during job execution. Assigning additional CPUs during execution incurs much higher cost and may not even be available leading to delays in job completion.
In summary, overly optimistic estimators such as median (P50) and mode can lead to underestimation of CPU needs, while conservative estimators (max and P75) can cause overestimation errors. Balancing between the two extremes is crucial to achieve efficient scheduling and resource allocation, as achieved by our proposed COSPiS approach.

\paragraph{Degree of Deadline Violation}
\begin{figure}[h!]
        \centering
            \centering
            \includegraphics[width=0.75\linewidth]{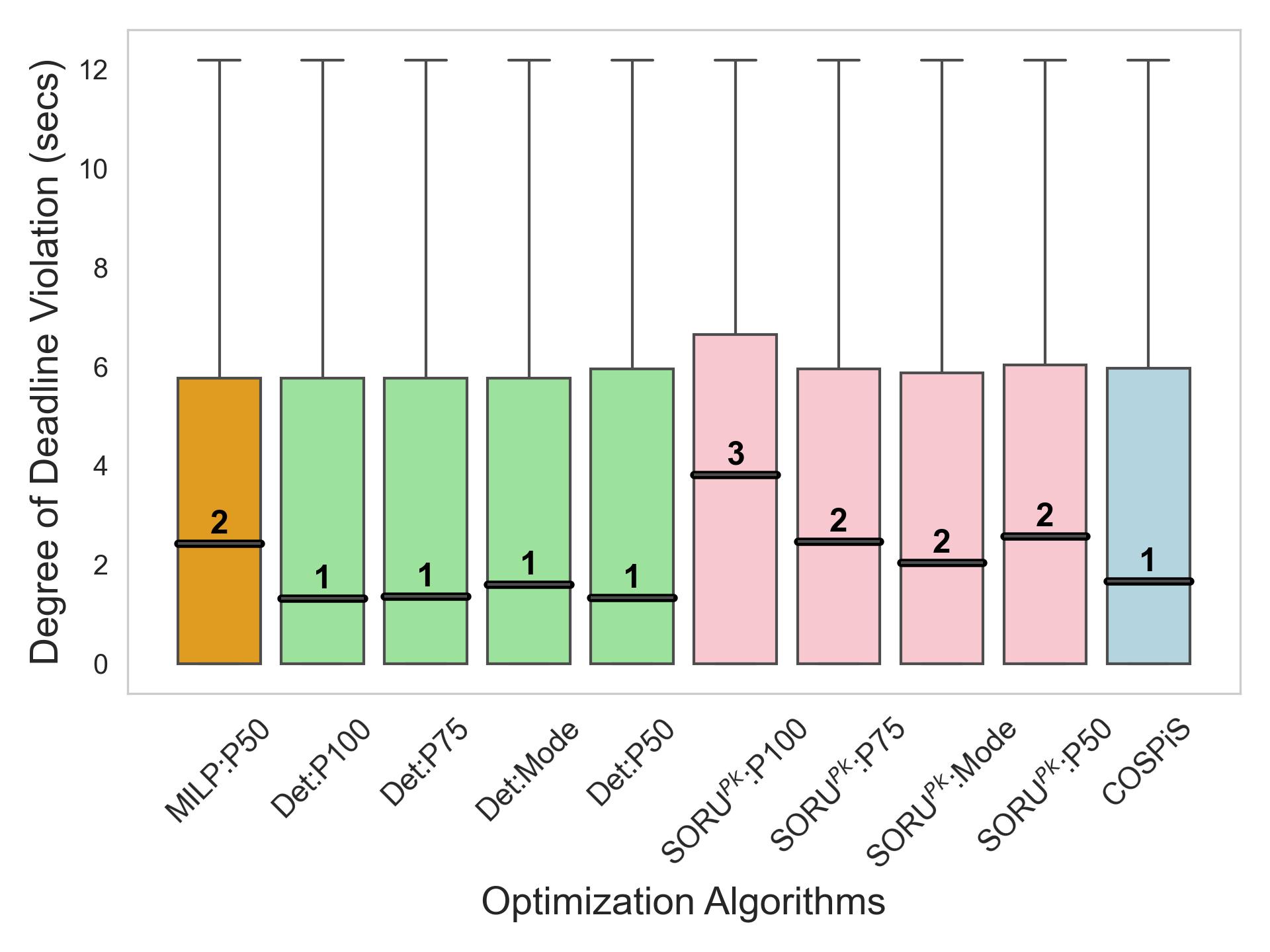}
            \caption{A comparison of the degree of deadline violation (lower $\rightarrow$ better) across a dataset of 44 COS problems from our organization, with each problem containing between 7 and 348 jobs. The plot shows four categories of approaches: (1) MILP with a Deterministic Estimator using a median (P50) estimator (orange), (2) Det: Constraint Programming with multiple Deterministic Estimators - max (P100), 75th percentile (P75),  median (P50) and mode (all shown in green), (3) SORU$^{Pk}$, as modified from \cite{varakantham2016proactive} to optimize peak CPU usage, employing estimators for CPU usage - max (P100), 75th percentile (P75),  median (P50) and mode (all shown in pink), and (4) COSPiS (Capacity Planning and Scheduling via Pair Sampling), our proposed algorithm (shown in blue). All approaches perform well with respect to the degree of deadline violation with violations restricted within few seconds. 
            }
         \label{fig:deadline_violation_all_approaches}
    \end{figure}

       Figure~\ref{fig:deadline_violation_all_approaches}  shows that all optimization algorithms demonstrate good performance regarding the degree of deadline violation, with delays never exceeding the order of a few seconds from their requested deadlines. 
       In the case of COSPiS, the selection of the right tolerance value (more details in section~\ref{sec:hyperparam_study}) assists in controlling the potential for deadline violation. For the deterministic estimator based approaches (Det and MILP), optimistic estimators have similar performance as conservative estimators because our dataset possesses a comfortable margin for job deadlines, preventing severe deadline violations. As a result, the deadline violations, when they occur are of order of just a few seconds.

    \subsubsection{Hyperparameter Study for COSPiS}
    \label{sec:hyperparam_study}

        \begin{figure}[!h]
    \centering

        \begin{subfigure}{0.49\textwidth}
        \centering
        \includegraphics[width=\linewidth]{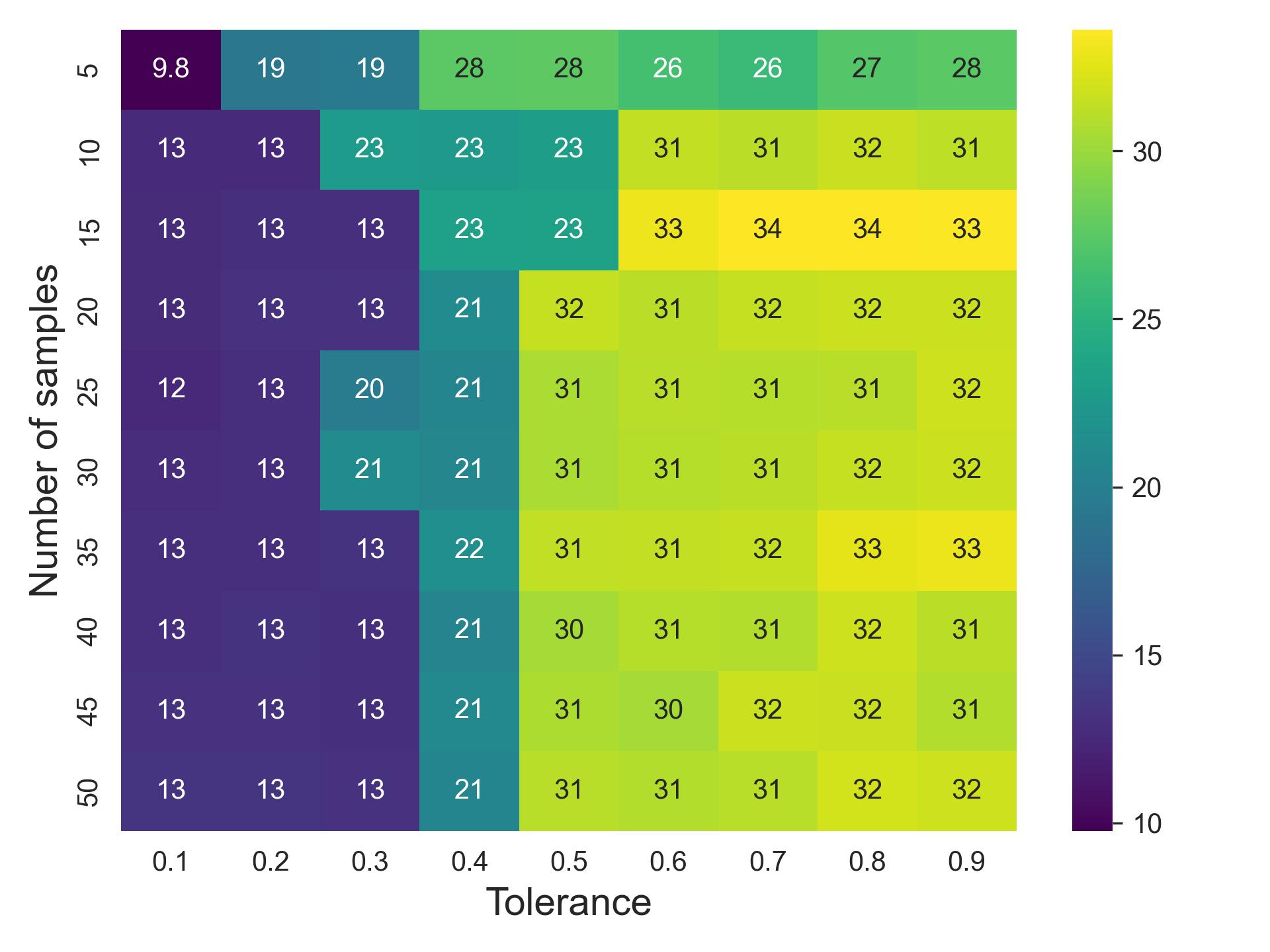}
        \caption{Avg. Peak Reduction (in \%tage) increases with increase in tolerance.}
        \label{fig:hyperparam_peak_reduction}
    \end{subfigure}
     \begin{subfigure}{0.49\textwidth}
        \centering
        \includegraphics[width=\linewidth]{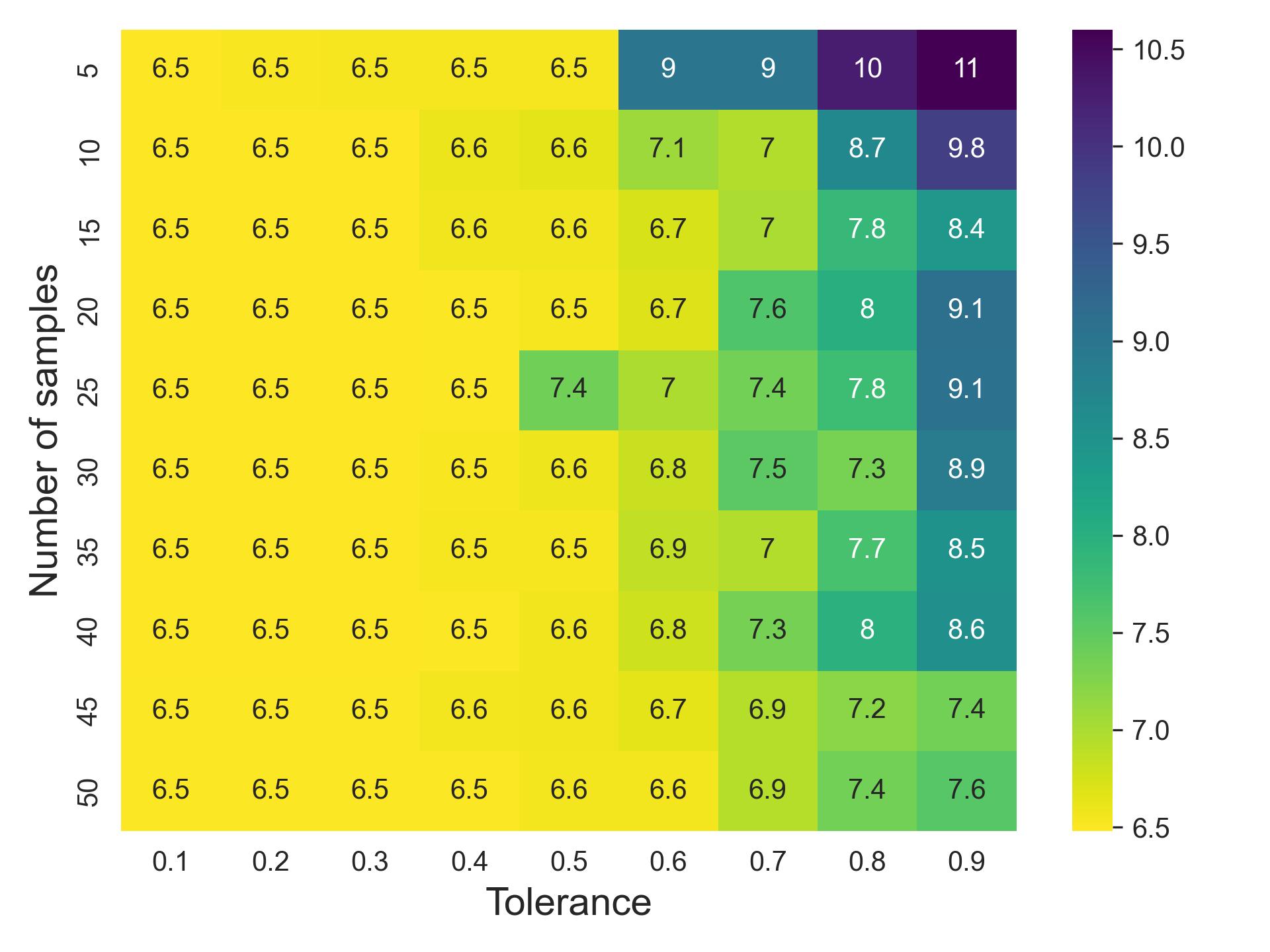}
        \caption{Avg. Degree of Deadline Violation (secs) decreases with increase in tolerance.}
        \label{fig:hyperparam_deadline_violation}
    \end{subfigure}
    \begin{subfigure}{0.49\textwidth}
        \centering
        \includegraphics[width=\linewidth]{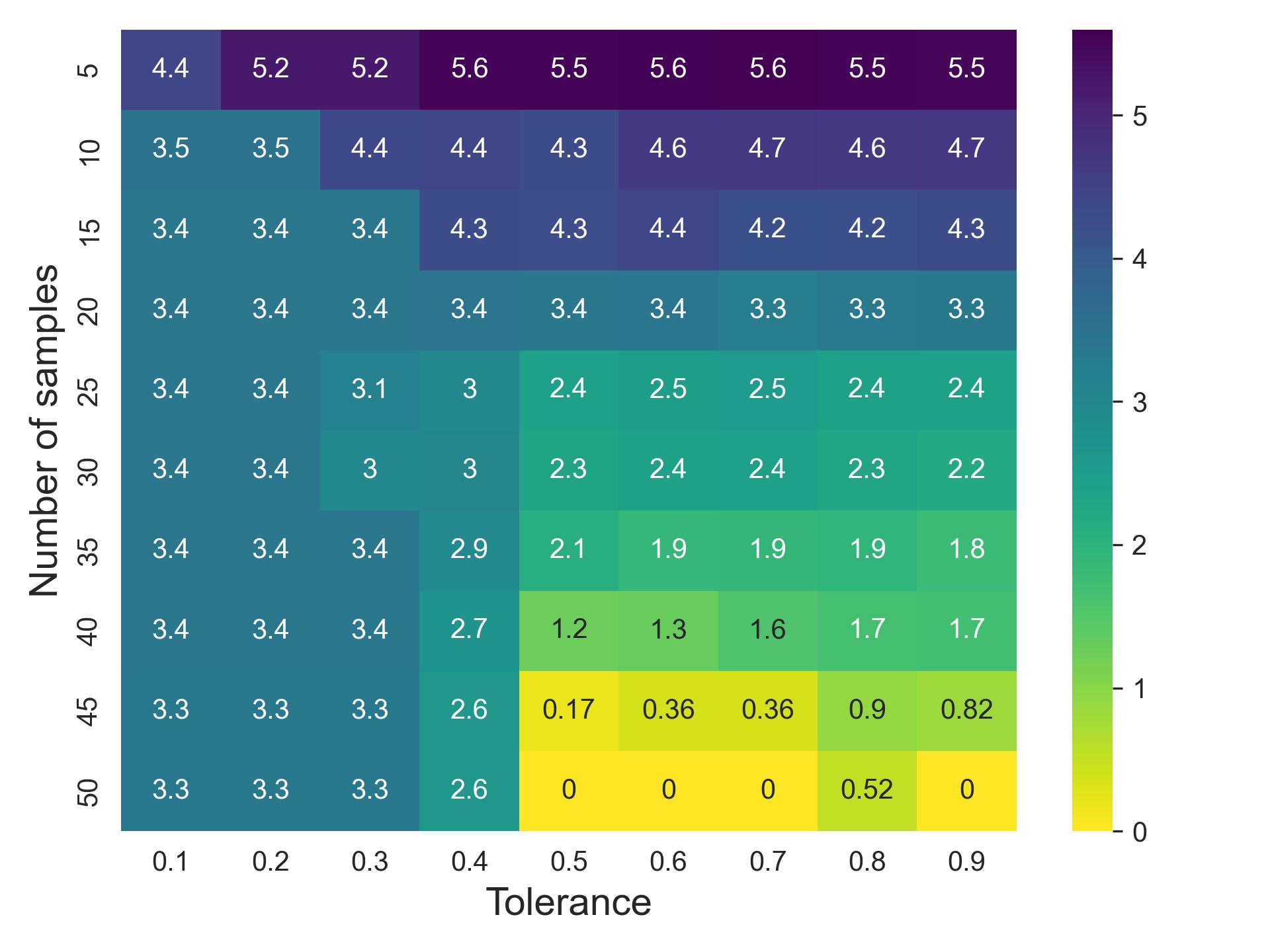}
        \caption{Capacity Under-Estimation Error (relative error in log scale) decreases with increase in the number of samples.}
        \label{fig:hyperparam_capacity_underestimation}
    \end{subfigure}
    \hfill
    \begin{subfigure}{0.49\textwidth}
        \centering
        \includegraphics[width=\linewidth]{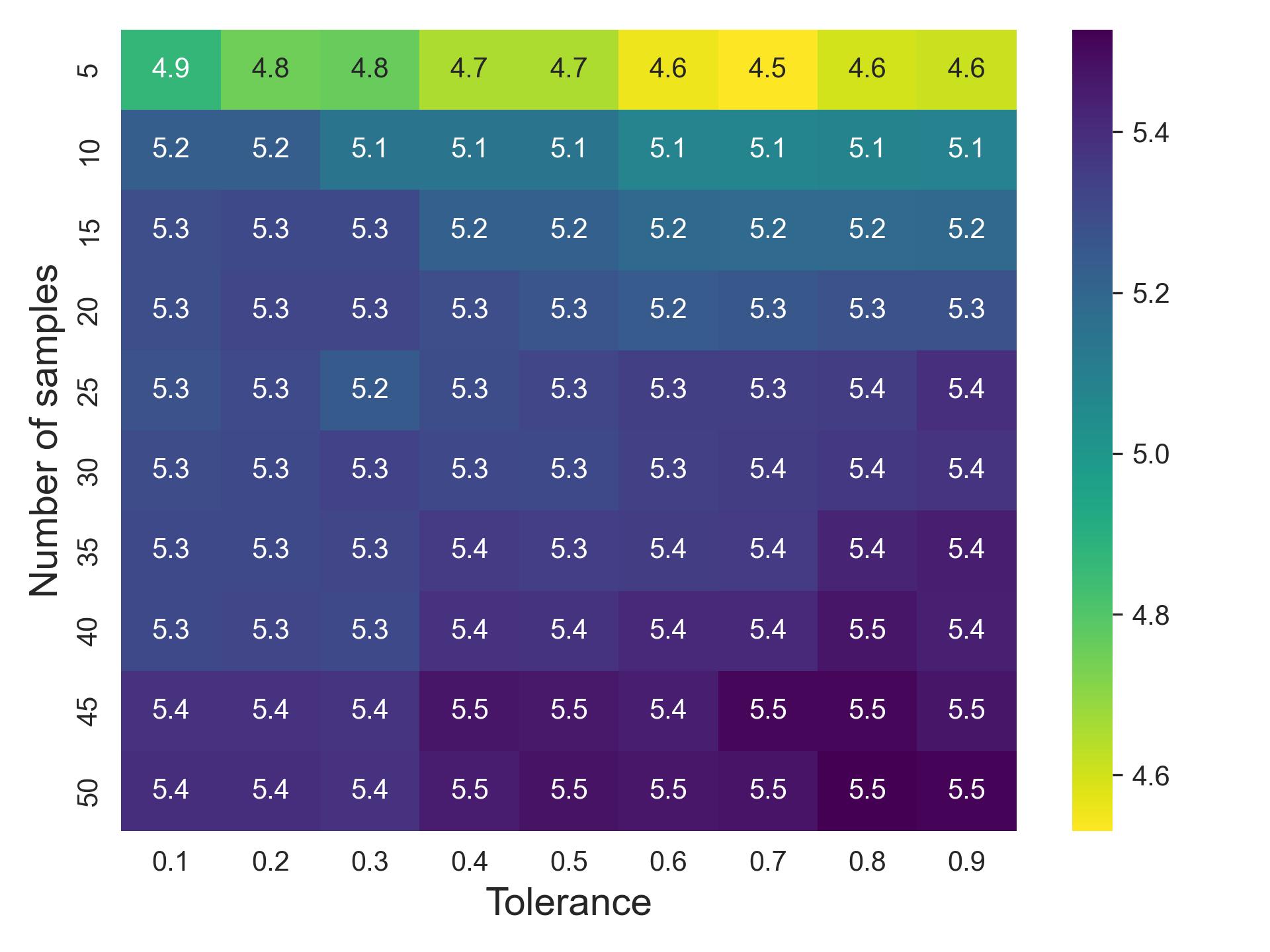}
        \caption{Capacity Over-Estimation Error (relative error in log scale) increases with increase in the number of samples.}
        \label{fig:hyperparam_capacity_overestimation}
    \end{subfigure}

    \caption{Hyperparameter study for our COSPiS approach showing four performance metrics (section~\ref{sec:performance_metrics}) for different values of (1) the number of samples taken from historical job duration and CPU usage data, and (2) tolerance: the maximum ratio of samples that may violate job deadline constraints. Brighter colors (yellow) indicate better performance. Based on this study, we set the number of samples to 25 and the tolerance to 0.4 for our experiments.
    }
    \label{fig:hyperparameter-tuning}
\end{figure}


Our proposed COSPiS approach has two hyperparameter: (1) the number of pair samples taken from historical job duration and CPU usage data, and (2) tolerance: the maximum ratio of samples that may violate job deadline constraints.
In order to effectively select the number of samples and tolerance, we conducted a hyperparameter study for a subset of four problems from our dataset. We varied the number of samples from 5 to 45 in steps of 5, and varied the tolerance from 0.1 to 0.9 in steps of 0.1. Figure \ref{fig:hyperparameter-tuning} shows our four performance metrics (section~\ref{sec:performance_metrics}) for all combinations of the hyperparameters.

We noticed an expected set of trade-offs as we varied the tolerance. With higher tolerance, COSPiS allows more samples to violate job deadlines, leading to a greater reduction in peak CPU usage (\ref{fig:hyperparam_peak_reduction}). However, as the tolerance value increased, the degree of deadline violations also increased (\ref{fig:hyperparam_deadline_violation}). We selected tolerance=0.4 which provides a good balance for overall efficiency in our CPU utilization, even as it pushed us closer to the edges of our deadlines.

On the other hand, when we increased the number of samples, we were faced with a different trade-off scenario. On the upside, increasing the samples led to a decrease in capacity under-estimation error (\ref{fig:hyperparam_capacity_underestimation}), thus improving our reliability to allocate adequate resources for the tasks. But at the same time, over-estimation error increased (\ref{fig:hyperparam_capacity_overestimation}), meaning that while we became better at ensuring that we were not lacking resources, we also risked allocating more resources than necessary, which could potentially lead to less efficient CPU usage overall. Based on these observations, we selected the combination of 25 samples and a tolerance of 0.4 as a suitable candidate because it balanced these competing metrics effectively. 

\subsubsection{Detailed Study of the Proposed Approaches}
\label{sec:detailed_exp_study}

In this section, we do an in-depth comparison of three of our proposed techniques, Det:P50, COSPiS, and MILP:P50 for COS problems of different sizes. This study aims to examine any any subtle variations that the large-scale tests may have potentially overlooked, thereby providing a detailed understanding of each method's strengths and challenges.

     \begin{figure*}[!t]
        \centering
       
        \begin{subfigure}{\linewidth}
            \centering
    \includegraphics[width=\linewidth]{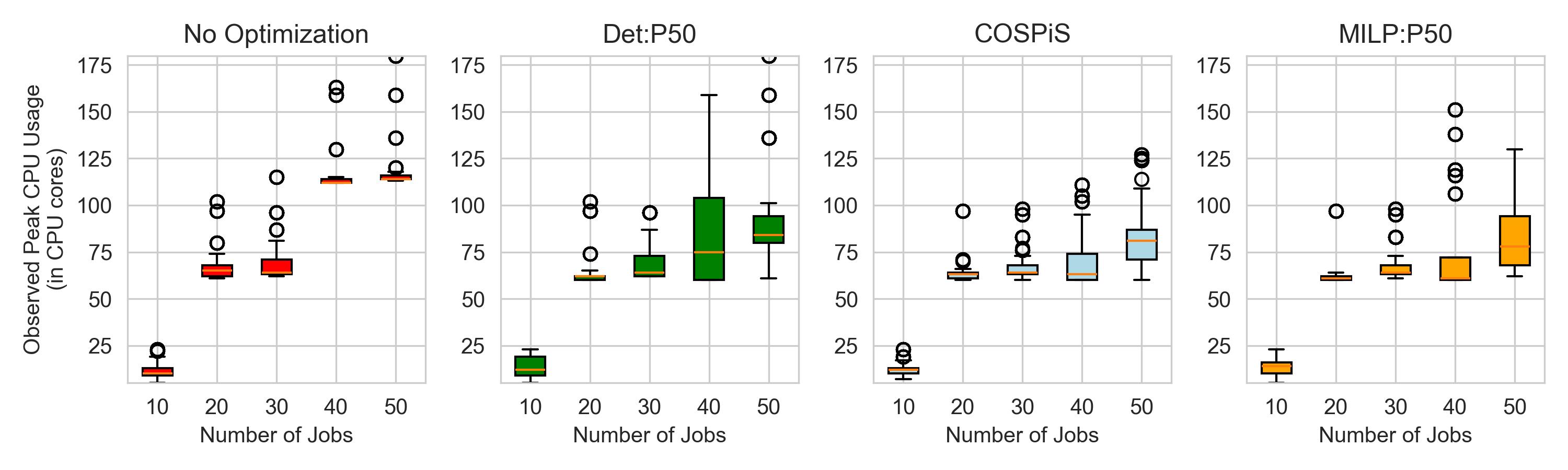}
            \caption{A subset of problems with number of jobs $\leq$ 50. The Det:P50 approach has an peak reduction of 7.79\% on an average, the COSPiS approach of 15.24\%, and the MILP:P50 approach of 13.29\% compared to no optimization (manual scheduling).}
            \label{fig:observed_peak_real}
        \end{subfigure}

                \begin{subfigure}{\linewidth}
            \centering
    \includegraphics[width=0.75\linewidth]{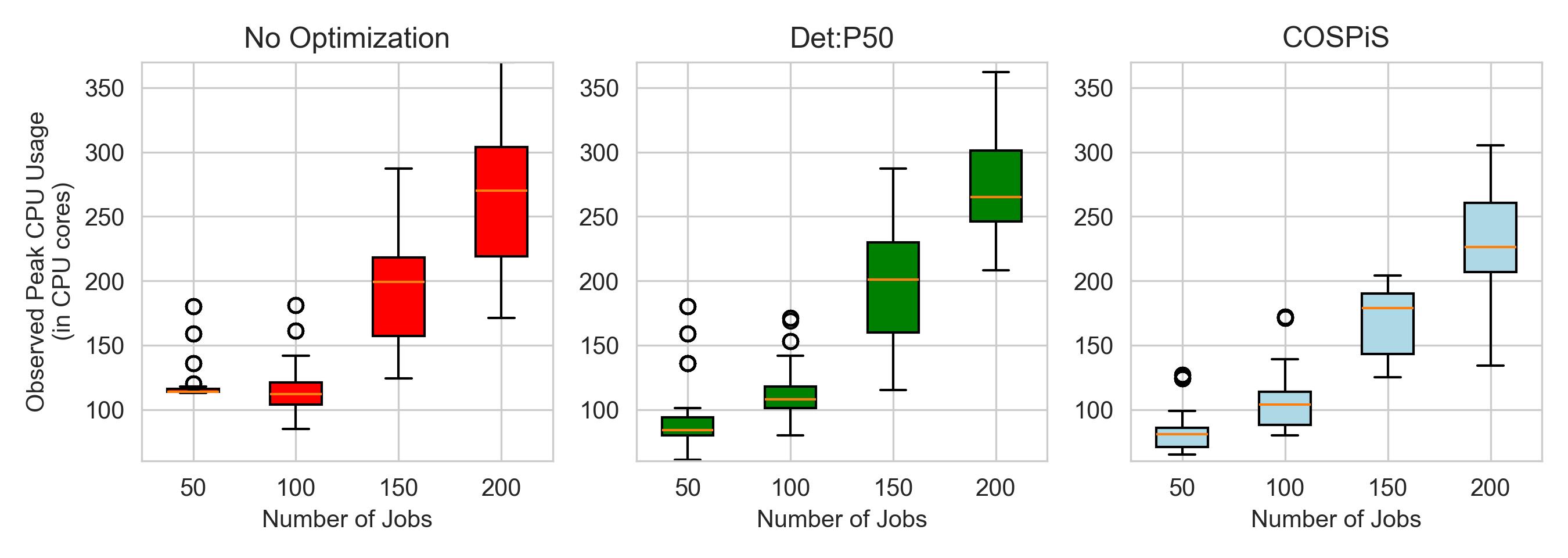}
            \caption{Scheduling for $50-200$ jobs. The Det:P50 approach has an average peak reduction of 6.12\% on an average, the COSPiS approach of 15.56\%. The MILP:P50 model was unable to find a feasible solution within 60 mins. }
            \label{fig:observed_peak_real_beyond_50}
        \end{subfigure}
        
        \caption{Comparison the observed peak CPU usage with respect to a subset of problems of different sizes (\#jobs) for three approaches: (1) no optimization (manual), (2) Det:P50, constraint programming approach with a median estimator, (3) COSPiS: a constraint programming approach with pair sampling, and (4) MILP:P50, a mixed-integer linear programming with a median estimator. Each box summarizes the result of 25 runs.}
        \label{fig:observed_peak}
    \end{figure*}

    \begin{figure*}[h!]
        \centering
        \begin{subfigure}{0.8\linewidth}
            \centering
            \includegraphics[width=\linewidth]{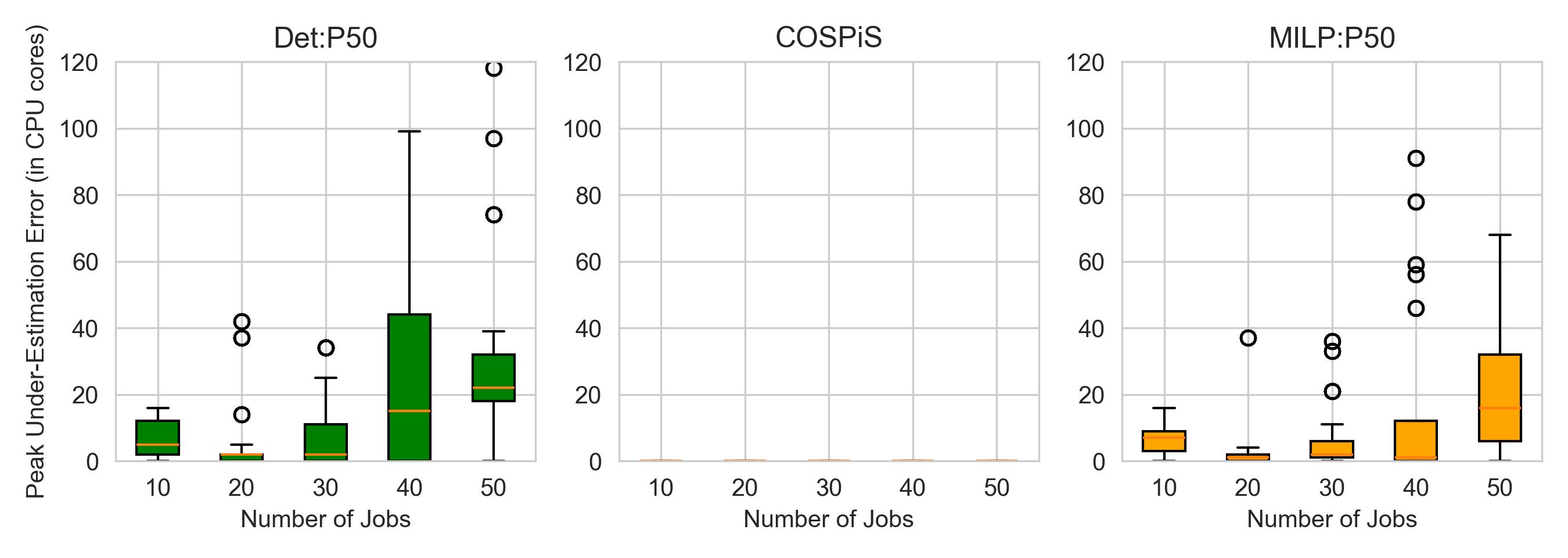}
            \caption{A comparison of the capacity under-estimation error. COSPiS approach performs the best with 0 capacity under-estimation error in all cases.}
            \label{fig:real_resource_violation}
        \end{subfigure}
        \hspace{0.2em}
        \begin{subfigure}{0.8\linewidth}
            \centering
            \includegraphics[width=\linewidth]{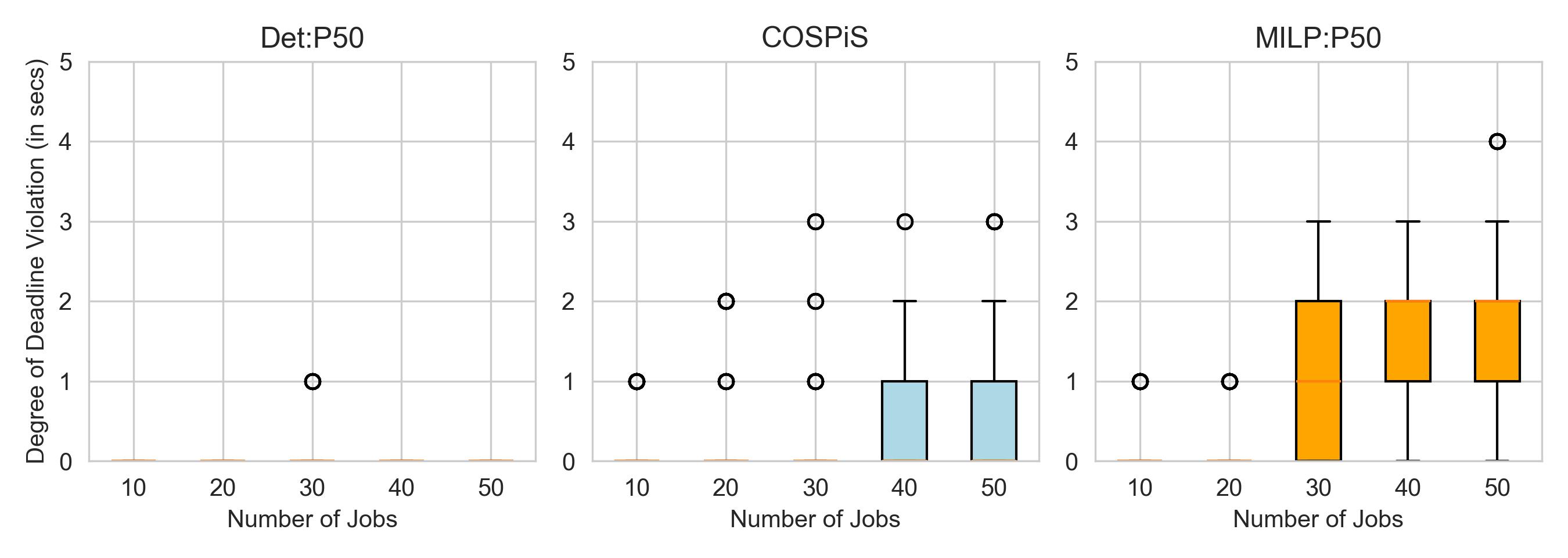}
            \caption{A comparison of the degree of deadline violation. All approaches have $\leq$ 4 secs delay in job completion which is insignificant.
            }
            \label{fig:real_deadline_violation}
        \end{subfigure}
        
        \caption{Results of quality of the two service metrics((a) and (b)) comparing the Det:P50 approach (left), the COSPiS approach (middle), and MILP:P50 (right). Each box summarizes the result of 25 runs.}
        \label{fig:experiments_real}
    \end{figure*}

        We start by comparing the observed peak CPU usage with the different approaches compared to manual scheduling (no optimization). Figure~\ref{fig:observed_peak} shows the observed peak for manual schedules and our three approaches with varying problem sizes from 10 to 200. Due to the stochastic nature of the jobs, each run (execution) of a schedule may result in a different peak.  
        
        The MILP:P50 model times out (at 1 hour) beyond 50 jobs. This is because it has more variables and constraints than the CP models. We observe in Figure~\ref{fig:model_MILP} that the MILP model has $O(n^2)$ more variables than the CP model in the form of $\delta^{1}_{ji}$, $\delta^{2}_{ji}$ and $res_{ji}$, and constraints for each of the additional variables. Figure~\ref{fig:observed_peak_real} compares the observed peak usage for up to 50 jobs and Figure~\ref{fig:observed_peak_real_beyond_50} shows the observed peak beyond 50 jobs. In both cases, the constraint programming with Sample Average Approximation (COSPiS) has the highest average peak reduction of 15.24\% in \ref{fig:observed_peak_real} and 15.56\% in \ref{fig:observed_peak_real_beyond_50}. In \ref{fig:observed_peak_real}, the schedule found by MILP:P50 performs second best with an average peak reduction of 13.29\%, however, it takes up to the order of ten times longer to compute.

        Figure~\ref{fig:experiments_real} summarizes the results in terms of the two quality of service metrics.
        Our COSPiS approach consistently predicts the peak CPU usage accurately even as the number of jobs increases, while the Det:P50 and MILP:P50 techniques degrade as we consider more jobs. This is the result of underestimation of the CPU usage in which the actual number of CPU cores used while executing the job is higher than the predicted value by the median estimator. 
        
         In terms of the degree of deadline violation metric (Figure~\ref{fig:real_deadline_violation}), we note that all approaches have the median violation amount of $\leq 2$ secs and a maximum of 4 secs. Even though our models are approximate, jobs are not delayed significantly. The Det:P50 and COSPiS have slightly better performance but the difference is not very significant and may be attributed to staggering delays in starting the jobs. 
        
        Overall, from  our experiments, we can conclude that the three approaches are successful in minimizing peak CPU usage. One advantage of using a median estimator is that it is simpler to model, in both constraint programming and MILP. However, it does very little in terms of handling uncertainty. It has a very high capacity under-estimation error, meaning, in the execution of the computed schedule, the CPU usage is more likely to be higher than the predicted $p$ and job deadline constraints are more likely to be violated. In other words, the median estimator approach is risky if there is a high degree of uncertainty and it leads to higher peak resource violations and deadline violations. In terms of providing both peak reduction and high quality of service to end users, the COSPiS approach is the most reliable model.
    

\begin{figure*}[h!]
        \centering
        \hfill
        \begin{subfigure}{\linewidth}
        \centering
            \includegraphics[width=0.7\linewidth]{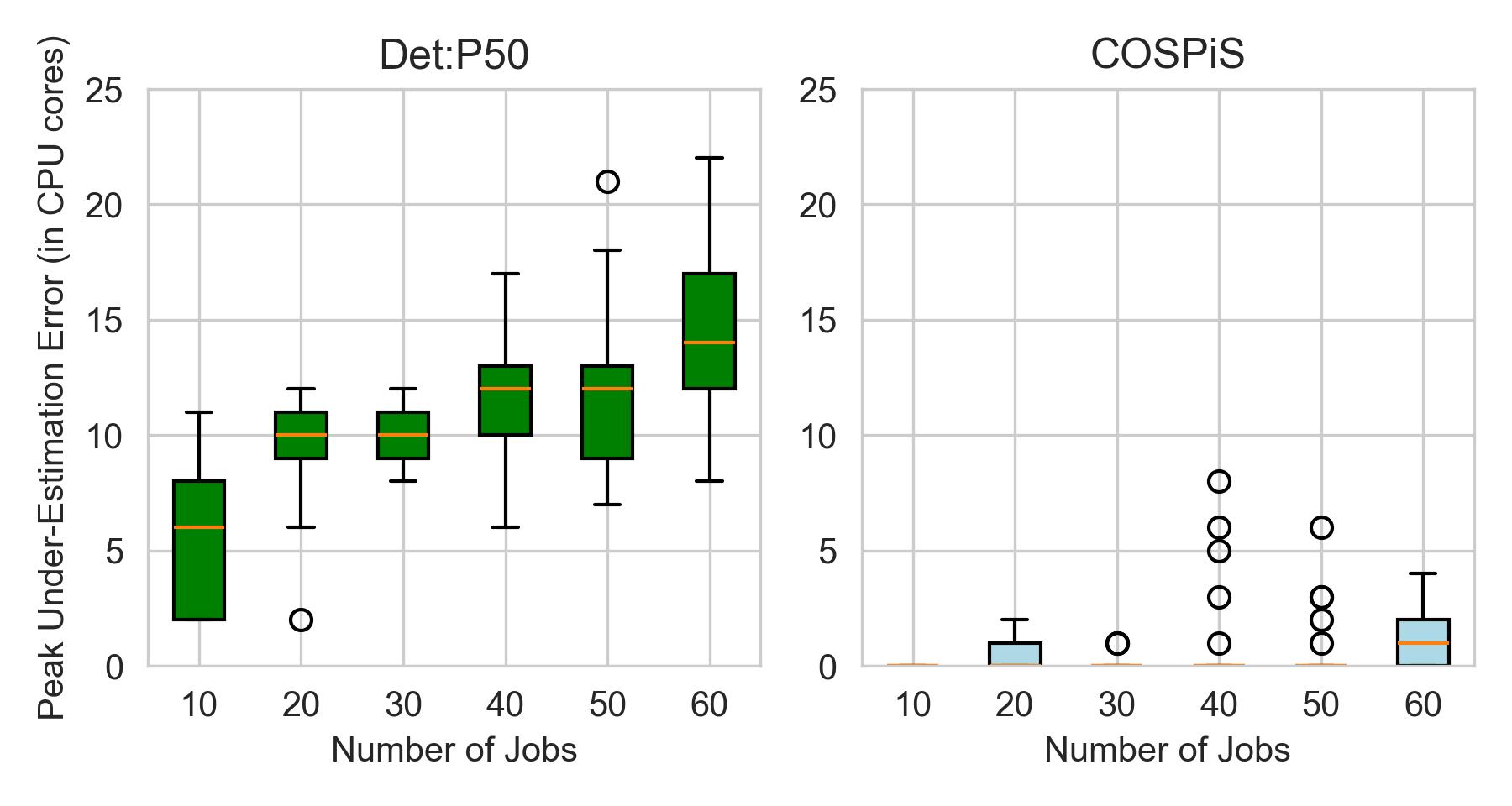}
            \caption{A comparison of the capacity under-estimation error of the Det:P50 (median estimator) approach and the COSPiS approach with 25 runs of each computed schedule by the respective approaches.}
            \label{fig:synthetic_resource_violation}
        \end{subfigure}
        
        \begin{subfigure}{\linewidth}
           \centering \includegraphics[width=0.7\linewidth]{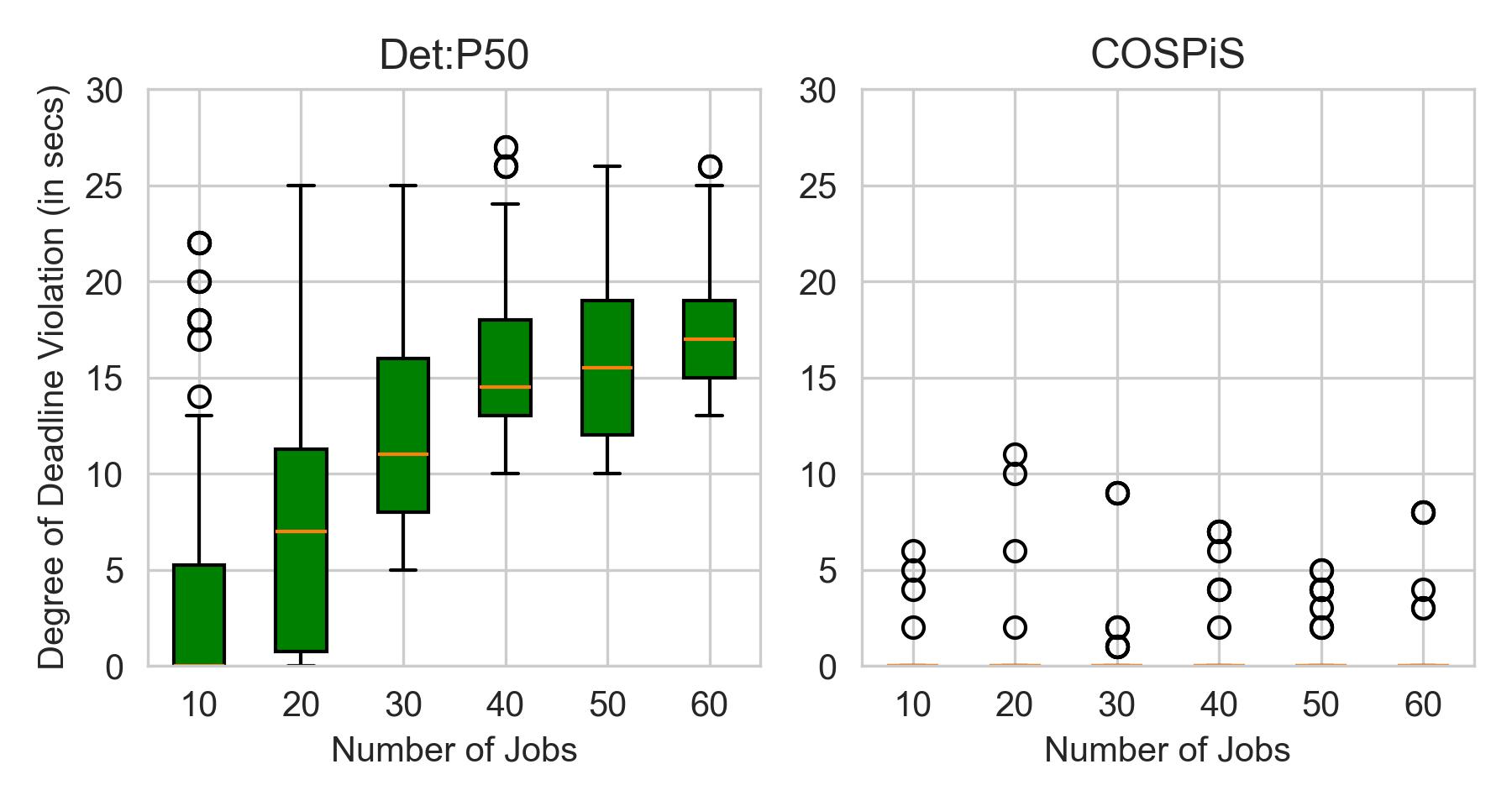}
            \caption{A comparison of the deadline violation of the Det:P50 approach and the COSPiS approach with 25 runs of each computed schedule by the respective approaches.}
            \label{fig:synthetic_deadline_violation}
        \end{subfigure}
        \caption{Results of our experiments with synthetic data comparing the two variants of our constraint programming approaches, median estimator (Det:P50) and Sample Average Approximation (COSPiS). 
        }
        \label{fig:experiments_synthetic}
    \end{figure*}

    \begin{figure*}[t]
        \centering
            \includegraphics[width=\linewidth]{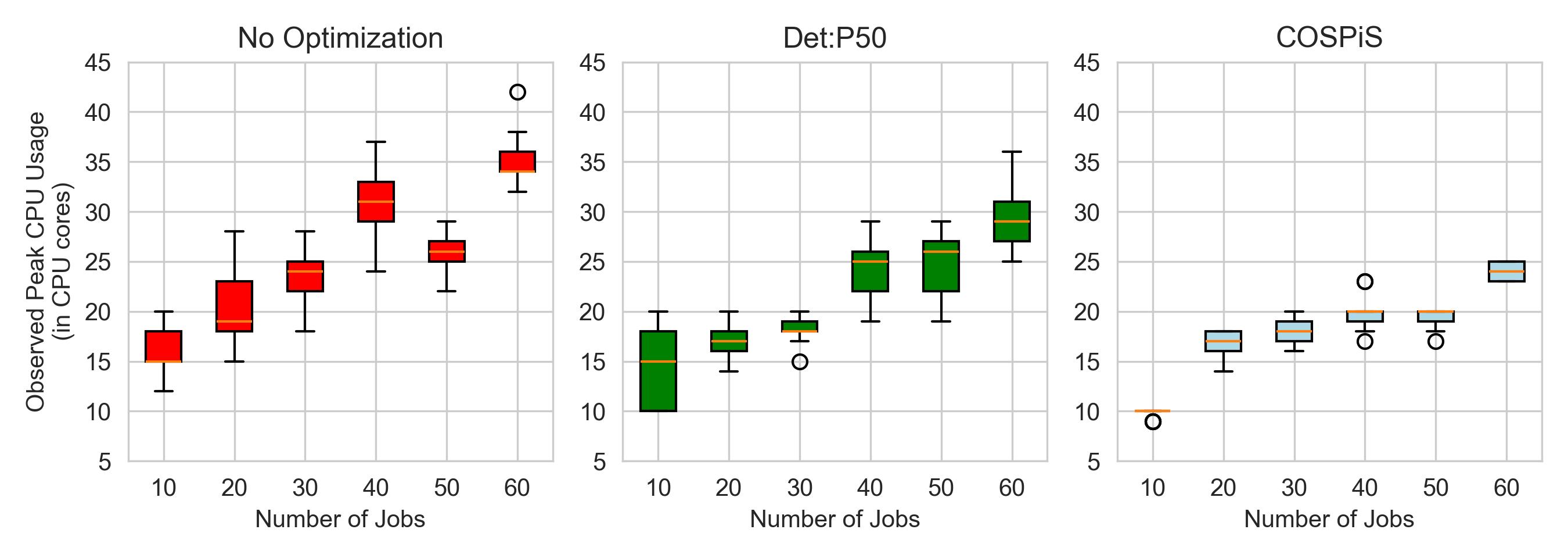}
        \caption{The observed peak CPU usage from executing jobs following three different schedules with synthetic data. (1) schedule with no optimization (manual scheduling), (2) Det: P50 (median estimator approach), and (3) COSPiS approach. Each box summarizes the result of 25 runs. On average, the median estimator approach has 15.65\% lower peak, the COSPiS approach has 28.87\% lower peak compared to no optimization.}
        \label{fig:observed_peak_synthetic}
    \end{figure*}

\clearpage   
    \subsection{Experiments with Synthetic Data}
        While the COSPiS approach did well on our organization's dataset, we wanted to verify whether the results hold for any type of data distribution in general. 
        To synthetically generate an input collection of jobs, we select the history of jobs' durations ($D$) and CPU usage ($R$)\footnote{Refer to Section~\ref{sec:problem_description} for definitions of $D$ and $R$.} from a predefined distribution. We choose the duration of a job to lie uniformly in the range [10, 30]\footnote{Unless mentioned otherwise, the unit of time is in seconds throughout the paper.}, and CPU usage uniformly in [5, 10]. Also for each job, $|D| = |R| = 50$. Furthermore, we have defined different COS problems $B_n$ where $n \in \{10, 20, 30, 40, 50, 60\}$. The makespan, $T$, for each scheduling problem $B_n$ is in the range $[500, 3000]$.  
        A job can depend on up to three other jobs.
        The requested start time ($q$) of a job is chosen uniformly within the makespan, and the flexibility ($f$) is randomly chosen from the set, \{20, 30, 80, 120\}. Finally, the deadline ($u$) is set to be $q + f + max(D)$. 
        
        \paragraph{Results} Figure~\ref{fig:experiments_synthetic} shows our results with the synthetic data. 
        In general, the predicted peak by COSPiS is higher than the median estimator. This is expected as the COSPiS approach incorporates several samples of each job, whereas, the median estimator takes an estimate which is more likely to be an optimistic approximation. 
        
        While considering the objective function, a lower peak is more favorable, but the robustness of such a solution is unknown.
        To test the robustness of such schedules, we executed the jobs according to the computed schedule by  the median estimator approach and calculated the capacity under-estimation errors. Figure~\ref{fig:observed_peak_synthetic} shows the observed peak CPU usage after job execution and  
        Figure~\ref{fig:synthetic_resource_violation} shows that the schedules generated using the median estimator approach always have higher capacity under-estimation error violations than COSPiS. The COSPiS approach consistently limits the degree of deadline violations even as the number of jobs increases, while the median estimator technique degrades as we consider more jobs. This is the result of underestimation of the CPU usage in which the actual usage of CPU cores while executing the job is higher than the predicted value by the median estimator. 
        
        We observe a similar pattern in the degree of deadline violation metric (Figure~\ref{fig:synthetic_deadline_violation}), as the COSPiS approach has lower deadline violation than the median estimator approach and the difference being more pronounced with the increase in the number of jobs in the scheduling problems.

\section{Job Scheduling and Execution Workflow}
    \label{sec:deployed_app}
    \begin{figure*}[!h]
        \centering
        \includegraphics[width=0.9\linewidth]{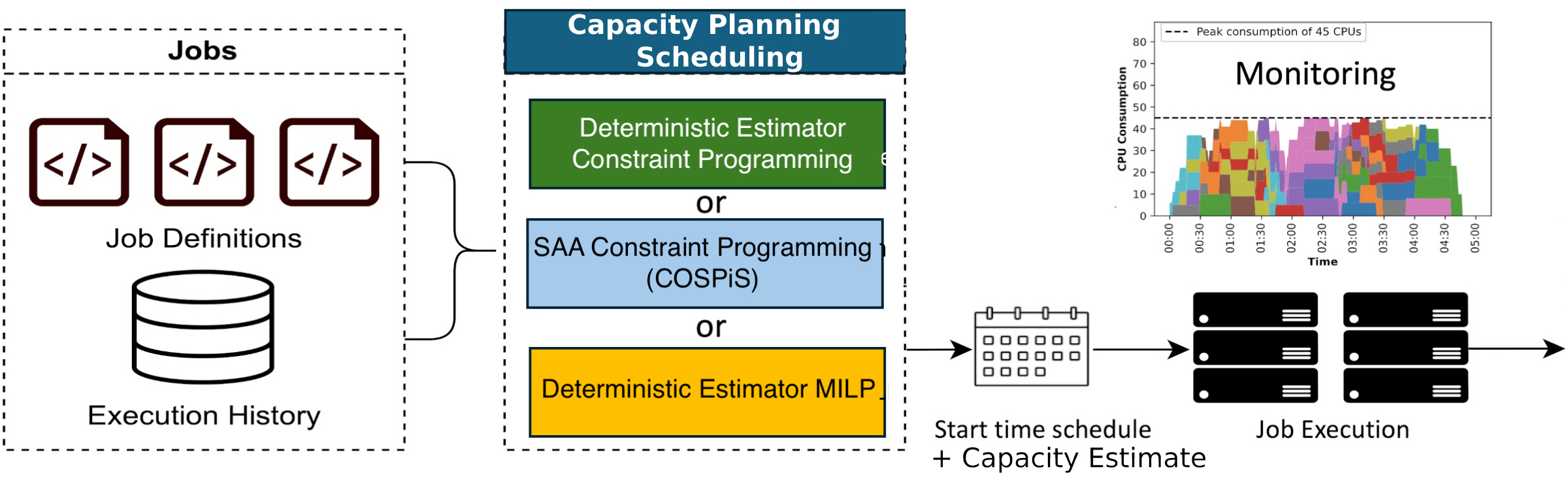}
        \caption{Job execution workflow showing the our three approximate optimization models (two constraint programming and one mixed integer linear programming) with the compute infrastructure of our organization.}
        \label{fig:pipeline}
    \end{figure*}
    
    We have an in-house risk estimation, pricing, and trade management platform designed to improve productivity  by offering our technologists, quantitative analysts, and risk managers a consistent, cross-asset portfolio of models, frameworks, and tools to use in building financial applications. In this section, we describe the job scheduling and execution workflow starting from the selection of job sets to their completion by their requested deadlines. 
    
    Our grid computing infrastructure is actively used by more than 4000 developers across various lines of businesses  making over 30,000 commits on a monthly basis. They run their business-critical workloads using a system $\mathcal{S}$. $\mathcal{S}$ allows users to specify various job parameters including start time and dependencies described in the paper, as well as supports distributed computation for calculation heavy workload – currently scheduling over 40,000 jobs in production on a daily basis in different partitions. Each partition is scheduled individually and has up to 400 jobs.

    As a pre-requisite to this work, we built a large data pipeline to warehouse job runtime characteristics and enrich it with the underlying job compute usage. This allowed us to build a coherent, temporal view of overall infrastructure usage patterns and, also have a profile compute requirements on a job-by-job basis. Currently, we have the ability to generate schedules on demand, tailored for the workload running on specific partitions (logical groupings of compute nodes for each line of business), with a focus on partitions with inconsistent compute usage patterns and high job deadline breaches.
The underlying grid-compute infrastructure is shared. If the CPU usage exceeds the predicted value on one of the partitions, then that partition ends up borrowing CPUs allocated to other partitions that are not in use at the moment. This leads to a shortage of CPUs in the immediate future and a ripple effect of deadline breaches. 

    \subsection{Architecture} Figure \ref{fig:pipeline} shows a schematic view of our job scheduling and execution workflow, composed of three main components; the job input queue, the scheduling engine, and an execution platform. The decision to choose between Det, COSPiS, MILP, or no optimization is made by the user. Each approach can be configured to consider uncertainty only for duration or only for CPU usage or both. One can run our job execution workflow without any optimization and kick-start the jobs at the requested start times by the user. 

    Once a schedule has been decided (by the user or the intelligent scheduling engine), the jobs are executed on the compute infrastructure honoring the schedule. However, a job may not start at its requested start time due to several issues. At the job execution level, each job is decomposed into several sessions, and sessions into tasks. A task is a basic execution unit on the compute infrastructure. There may be some lags between sessions and tasks of a job due to unexpected events, such as a sudden arrival of higher priority jobs, downtime of cores for repair/maintenance, etc. If a job is a pre-requisite for other jobs, this may cause a staggered delay effect. 

    \paragraph{Job Monitoring.} Once a job is submitted for execution with a particular start time, it enters the execution queue. We have a real-time job monitoring dashboard that shows the current status \texttt{\{waiting, running, succeeded, failed\}} of all jobs. While a job is running, the user can stop the execution of the job if they feel it is necessary to do so, and new jobs may arrive online. 
    The monitoring process also shows the CPU usage at a single time point. Once a job is completed, the status is informed to the user by their chosen mode of communication. 
    

\section{Discussion}

    Although we focus on one category of resource (CPUs) in this paper, we can generalize this approach to any number of resources. Our proposed COSPiS approach has two hyperparameters: the number of samples and tolerance. Currently, we tune these hyperparameters empirically by checking the performance for a wide range of values. Some of the scheduling problem instances may be infeasible to solve for a specific value of tolerance. The deterministic estimator-based approaches (Det, MILP, SORU) also run into infeasible solutions when estimates are conservative. 
    In the current setup, we manually re-run the scheduler with different values of tolerance in such cases. If still no solution is found, the problem follows the manually suggested start times. To improve upon this manual strategy, we can integrate online hyperparameter tuning algorithms to find feasible solutions without human interference.

    The scalability of our approach with respect to the number of samples and the number of jobs is 100 samples for up to 400 jobs for the constraint programming approaches. For the MILP model, the solver times out at 1 hour for more than 50 jobs. We also developed a robust stochastic optimization model with uncertainty sets \citep{chen2020robust}. However, it was quite slow (could schedule only up to 10 jobs in one hour) compared to the other models and we chose not to include it in the paper. To do smarter and more efficient sampling, one can develop a distributionally robust optimization model of the stochastic scheduling problem.

    
    To test the robustness of our approach, one can experiment with different types of deadline distributions, varying makespan, different duration, and CPU usage distributions.
    Another objective we are interested to integrate with our model is to minimize the movement of jobs from their requested start times and distribute the jobs uniformly across different resources, and start times throughout the day.

    In this paper, we mainly focus on the problem of scheduling under uncertainty and don’t consider any fairness metrics associated with the jobs. For example, should individuals that submit only a few tasks should be more likely to have their task completed in time, compared to those who submit many? This would be an interesting feature to add in the future and useful if jobs were submitted by competing teams. In the current model, jobs have priority, and high-priority/critical jobs can be indicated by setting a low flexibility($f$) value.


\section{Conclusion} \label{sec:conclusion}
    
   Capacity planning and job scheduling is a critical problem to save environmental footprint and financial costs. The impact is significant for hybrid grid-compute structures which have thousands of jobs running every day. 
    We have proposed three approximate solutions, (a)  constraint programming  with a deterministic estimator (Det), (b) constraint programming with paired Sample Average Approximation (COSPiS), and, (c) mixed integer linear programming with a deterministic estimator (MILP). The objective is to minimize the peak CPU usage and provide an accurate capacity estimate while considering the uncertainty {\em both} in the jobs' running time (duration) and CPU usage.
    An evaluation of our proposed approaches provides strong evidence that we successfully deliver on our objective of capacity planning while ensuring low capacity under-estimation error and deadline violations. Using our COSPiS approach, we see an estimated peak CPU usage reduction of up to 41.6\%, compared to manual scheduling.
    Future work will include incorporating more sophisticated sampling techniques, and developing a distributed robust optimization model.

\section*{Statements and Declarations}

\noindent \textbf{Acknowledgements.} The authors would like to acknowledge Alberto Pozanco, Rui Silva, and Daniel Borrajo for their helpful suggestions and comments on this work. This paper was prepared for informational purposes in part by the Artificial Intelligence Research Group of JPMorgan Chase \& Co and its affiliates ("J.P. Morgan"), and is not a product of the Research Department of J.P. Morgan. J.P. Morgan makes no representation and warranty whatsoever and disclaims all liability, for the completeness, accuracy or reliability of the information contained herein. This document is not intended as investment research or investment advice, or a recommendation, offer or solicitation for the purchase or sale of any security, financial instrument, financial product or service, or to be used in any way for evaluating the merits of participating in any transaction, and shall not constitute a solicitation under any jurisdiction or to any person, if such solicitation under such jurisdiction or to such person would be unlawful.

\bibliography{references}

\appendix

\section*{Table of Symbols}

\begin{tabular}{|c|p{10cm}|}
    \hline
    \textbf{Symbol} & \textbf{Description} \\
    \hline
    COS & Capacity Optimization and Scheduling \\
    \hline
    COSPiS & Capacity Optimization and Scheduling via Paired Sampling \\
    \hline 
    Det & Deterministic Estimator-based Constraint Programming Approach \\
    \hline
    MILP & Mixed-Integer Linear Programming \\
    \hline 
    $b$ & A job represented as a tuple $(q, f, u, D, J, R)$. \\
    \hline
    $q$ & The requested start time of job $b$. \\
    \hline
    $f$ & Flexibility measure indicating the maximum delay allowed for the start of job $b$ after its requested start time $q$. \\
    \hline
    $u$ & The latest completion time (deadline) of job $b$. \\
    \hline
    $D$ & A list of recorded durations or running times from historic data of job $b$'s previous executions. \\
    \hline
    $J$ & The set of jobs that job $b$ depends on; job $b$ can only start once all jobs in $J$ have been completed. \\
    \hline
    $R$ & The history of the number of CPU cores utilized by job $b$. \\
    \hline
    $B_n$ & The set of $n$ jobs, each represented as $b_j$. \\
    \hline
    $S_n$ & A schedule for $n$ jobs, represented as $(s_1, s_2, \dots, s_n)$, where $s_j$ is the scheduled start time of job $b_j$. \\
    \hline
   $T$ & The maximum timespan (makespan) within which all jobs need to run. \\
    \hline
    $S^*_n$ & The optimal start-time schedule for $B_n$ within a makespan of $T$. \\
    \hline
    $\{s_j\}^n_{j=1}$ & A set of integer variables where $s_j$ indicates the start time of job $b_j \in B_n$. \\
    \hline
    

    \textbf{Symbol} & \textbf{Description} \\
    \hline
    $p$ & An integer variable indicating the maximum (peak) number of CPU cores used across all jobs at any time $t \in T$. \\
    \hline
    $\hat{b}_j$ & A job $b_j$ mapped using a deterministic estimator function $\boldsymbol{\mathit{f^{est}}}$. \\
    \hline
    $\hat{d}_j$, $\hat{r}_j$ & Estimations of the duration and CPU usage of job $b_j$, respectively. \\
    \hline
    $\mathbf{X}$, $\mathbf{Y}$ & Sets of job runtime intervals and resource usages for $n$ jobs, used in the cumulative constraint. \\
    \hline
    $f^{est}$ & Estimator function \\
    \hline
    $K$ & Hyperparameter of COSPiS (number of pair samples) \\
    \hline 
    $\alpha$ & Hyperparameter of COSPiS (tolerance of job deadline violations) \\
    \hline 
    \end{tabular}

\end{document}